\documentclass[fleqn,usenatbib]{mnras}

\usepackage{newtxtext,newtxmath}

\usepackage[T1]{fontenc}
\usepackage{subcaption}
\usepackage[utf8]{inputenc}
\usepackage{tabularx}

\usepackage{graphicx}	
\usepackage{amsmath}	

\title[Resistive MHD simulations TRAPPIST-1e]{Resistive MHD Simulations of Stellar Wind–Magnetosphere Coupling in TRAPPIST-1e}

\author[Gonz\'alez-Avil\'es et al.]{
J. J. Gonz\'alez-Avil\'es,$^{1}$\thanks{E-mail: jgonzaleza@enesmorelia.unam.mx}
N. Baltazar P\'erez-Negr\'on,$^{1}$ and
A. Segura$^{2}$
\\
$^{1}$ Escuela Nacional de Estudios Superiores (ENES) Unidad Morelia, Universidad Nacional Aut\'{o}noma de M\'{e}xico, 58190 Morelia, Michoac\'{a}n, M\'{e}xico\\
$^{2}$ Instituto de Ciencias Nucleares, Universidad Nacional Aut\'onoma de M\'exico, A. P. 70-543 04510 D.F., Mexico 
}

\date{Accepted XXX. Received YYY; in original form ZZZ}

\pubyear{\the\year{}}

\begin{document}
\label{firstpage}
\pagerange{\pageref{firstpage}--\pageref{lastpage}}
\maketitle

\begin{abstract}

Close-in terrestrial exoplanets around M dwarfs reside in dense, magnetized winds, where non-ideal plasma coupling can strongly affect how electromagnetic energy is redistributed within the dayside interaction region. We present three-dimensional resistive magnetohydrodynamic simulations of the TRAPPIST-1 wind interacting with a dipolar TRAPPIST-1e magnetosphere for three stellar-wind forcing cases and four prescribed magnetic diffusivities, $\eta=(0,\ 538.018,\ 5.38018\times10^{8},\ 5.38018\times10^{12})$ cm$^{2}$ s$^{-1}$. Energy transport is diagnosed using maps of the total energy density, the magnitude of the total Poynting flux, and the divergence of the total Poynting flux. We further estimate a radio-power proxy from the volume integral of $\nabla\cdot \mathbf{S}_{\rm total}$ over the dayside bow-shock and magnetopause layers. Across all cases, increasing prescribed $\eta$ broadens the coupling layer and shifts the dominant energy-conversion regions from thin, patchy boundary arcs to thicker, more spatially extended structures, with an increasing relative contribution from the magnetopause. The inferred radio-power proxy increases by several orders of magnitude across the explored scan. However, because the estimated numerical magnetic diffusivity in the strongest-gradient regions is $\eta_{\rm num}\sim10^{15}$--$10^{16}$ cm$^{2}$ s$^{-1}$, the present $\eta$ scan is best interpreted as a controlled sensitivity study rather than as a direct constraint on the physical diffusivity of the TRAPPIST-1e environment. For the adopted planetary fields ($B_{\rm eq}=0.32$--$1.28$ G), the maximum cyclotron frequencies are $\nu_{c,\max}\approx1.8$--$7.2$ MHz, below the ground-based window, implying that meaningful radio constraints on TRAPPIST-1e magnetism will require space-based observations below 10 MHz or substantially stronger planetary fields than those assumed here.
\end{abstract}




\section{Introduction}
\label{sec:Intro}

The discovery of the first exoplanet represented a breakthrough in astrophysics, carried out in an era of rapid advances in our understanding of planetary systems beyond the Solar System \citep{Schneider2011}. Space-based missions such as Kepler \citep{Haas_et_al_2010} and the Transiting Exoplanet Survey Satellite \citep[TESS;][]{Ricker_et_al_2015} have collected large statistical samples of exoplanets, providing detailed insights into their masses, radii, and orbital properties. Among the most commonly detected are hot Jupiters, which are a class of gas giant exoplanets in close orbits around their host stars. These planets often reside within the stellar Alfv\'en radius, where the interaction between the stellar wind and the planetary magnetic fields can be strong \citep[e.g.][]{Shkolnik_et_al_2003, Cohen_et_al_2018}. As such, they serve as valuable systems for investigating magnetic star–planet interactions and associated dynamical processes.

Although considerable progress has been made in the detection and characterization of exoplanets, direct measurements of their magnetic fields remain inaccessible with current observational capabilities \citep{Strugarek-etal2025, strugarek-Shkolnik2025}. However, magnetic fields play a key role in the interaction with the host star and are linked to the planet internal structure and its atmospheric evolution \citep{Perryman2018}. In particular, for close-in exoplanets, magnetic fields can modulate the coupling between the stellar wind and the planetary magnetosphere, potentially giving rise to observable signatures in ultraviolet (UV), infrared, and radio \citep{Strugarek-etal2025, strugarek-Shkolnik2025}. These emissions offer indirect diagnostics of the magnetic environment and the underlying interaction processes.

A central mechanism in these systems might be the reconnection between the planetary magnetosphere and the magnetic field of the stellar wind. Under super-Alfv\'enic conditions, reconnection near the magnetopause can accelerate electrons along magnetic field lines towards the star. The resulting particle precipitation may generate cyclotron radio emission, providing a potentially observable signature of magnetic activity in exoplanetary systems. The interaction between main-sequence M-type stars (M dwarfs or red dwarfs) and their orbiting exoplanets is of particular interest for several reasons. M dwarfs possess intense magnetic fields, ranging from approximately 1 to 7 kG \citep{Kochukhov2021}. Hot-Jupiters around M dwarfs are scarce compared to other stellar types \citep[e.g.][]{Ganetal2023}; instead, they have high occurrence rates of planets with radii or masses consistent with a rocky composition \citep[e.g.][]{Sabottaetal2021,MentCharbonneau2023}. Additionally, the detection of terrestrial exoplanets in the habitable zone of M dwarfs is possible with current exoplanet detection techniques.

A variety of numerical models has been used to investigate the interaction between stellar winds and planetary magnetospheres. Ideal magnetohydrodynamic (MHD) simulations with the PLUTO code \citep{Mignone_et_al_2007} have been widely applied to different planetary environments, including studies of Mercury’s global magnetospheric configuration \citep{2015P&SS..119..264V,2016P&SS..129...74V,2016P&SS..120...78V,2016P&SS..122...46V}, Earth’s response to coronal mass ejections (CMEs) \citep{2022A&A...659A..10V,2023MNRAS.525.4008V}, and the propagation of slow-mode structures in Mercury’s magnetosphere \citep{2016P&SS..125...80V,2022A&A...659A..10V}. Other studies include modeling magnetospheric radio emissions in both solar system and exoplanetary settings \citep{2016A&A...595A..69V,2018A&A...616A.182V,2022SpWea..2003164V,2023ApJ...959L..13M}, evaluating space weather conditions around M dwarfs such as the Proxima~b system \citep{Pena-Monino_et_al_2024}, and preliminarily predicting the habitability and radio emission of TRAPPIST-1e \citep{ Wang_et_al_2025}. Non-ideal MHD approaches have also been employed, incorporating resistivity, Hall effects, radiative cooling, viscosity, and thermal conduction in simulations of star–planet wind–magnetosphere interactions \citep{2004ApJ...602L..53I,10.1093/mnras/staf707}, stellar and young-star magnetospheres and winds \citep{Gu_2009,Zanni_Ferreria_2013}, and solar-system magnetospheres \citep{https://doi.org/10.1029/2022JA030990,https://doi.org/10.1029/2022JA030280}.

Although previous numerical MHD studies have investigated stellar wind–magnetosphere interactions, the role of magnetic diffusivity in close-in M-dwarf systems such as TRAPPIST-1e remains unexplored. TRAPPIST-1e is a particularly well-motivated target for global magnetospheric modeling because it is a terrestrial planet (M = 0.692 M$_\oplus$, R=0.92 R$_\oplus$), orbiting its host star at 0.02925 au with a period of 6.1 days \citep{2016Natur.533..221G,2017Natur.542..456G,agol-etal-2021}. TRAPPIST-1e is located in the habitable zone and, according to several numerical models, may be capable of maintaining surface liquid water and a CO$_2$-dominated atmosphere \citep{Wolf2017,Tuerbetetal2018,Krissansen-Totton_Fortney2022,Sergeevetal2022}. These characteristics render TRAPPIST-1e a compelling candidate for potential planetary habitability, where atmospheric retention and star–planet interactions are central to habitability assessments \citep{Giallucaetal2024, Roettenbacher_Kane2017, Bourrieretal2017}. TRAPPIST-1 itself is an ultra-cool M8V dwarf star, with a mass of 0.08 M$_\odot$ and a radius of $R_\star = 0.114$~R$_\odot$, located at a distance of about 12 parsecs (pc). It hosts at least seven terrestrial planets in compact orbits ranging from 0.01 to 0.063 AU, with orbital periods between 1.5 and 20 days \citep{2016Natur.533..221G,2017Natur.542..456G}, making it one of the most compact planetary systems known to date. In addition, the TRAPPIST-1 system is a nearby benchmark M-dwarf laboratory that is frequently targeted for indirect magnetospheric diagnostics, including radio observations \citep[e.g.][]{Garraffo_2017,Wang_et_al_2025}.

Building on the framework of \citet{2018A&A...616A.182V}, we perform three-dimensional resistive MHD simulations with the PLUTO code to model the interaction between the TRAPPIST-1 stellar wind and an intrinsically magnetized TRAPPIST-1e. The primary objective of this work is to study how magnetic diffusivity modifies the large-scale interaction geometry and magnetospheric dynamics. As a secondary objective, we assess how magnetic diffusivity affects the radio emission driven by stellar wind–magnetosphere coupling, thereby informing the prospects for detectability with current and forthcoming radio facilities.

The paper is organized as follows. In Section~\ref{sec:Model_and_methods}, we describe the governing MHD equations, the numerical setup and methods, the initial and boundary conditions, and the simulation cases. Section~\ref{sec:Results_numerical_simulations} presents the results of the numerical simulations for the three cases and across the explored magnetic diffusivity values, including equatorial and meridional maps of key physical variables, as well as estimates of the radio emission based on the integrated power proxy computed from the volume integral of the divergence of the total Poynting flux. Finally, in Section~\ref{Discussion_and_conclusions}, we present the discussion and conclusions.

\section{Model and methods}
\label{sec:Model_and_methods}

To investigate the interaction between the stellar wind of TRAPPIST-1 and the intrinsic magnetosphere of TRAPPIST-1e, we employ the non-ideal magnetohydrodynamics (MHD) equations, incorporating magnetic diffusivity. These equations, derived from the conservation of mass, momentum density, and total energy density, together with the induction equation for a fully ionized plasma, can be expressed in cgs units as

\begin{equation}
\frac{\partial\rho}{\partial t} + \nabla\cdot(\rho{\bf v}) = 0,
\label{mass_cons}
\end{equation}

\begin{equation}
\frac{\partial{\bf m}}{\partial t} + \nabla\cdot\left[{\bf m}{\bf v}-\frac{{\bf B}{\bf B}}{4\pi} + I\left(p+\frac{{\bf B}^{2}}{8\pi}\right)\right]^{T} = 0,
\label{mom_cons}
\end{equation}

\begin{equation}
\frac{\partial E_t}{\partial t}
+\nabla\cdot\left[
\left(E_t+p_t\right)\mathbf{v}
-\frac{\left(\mathbf{v}\cdot\mathbf{B}\right)\mathbf{B}}{4\pi}
\right] =
\nabla\cdot\left(\frac{\eta}{c}~\mathbf{J}\times\mathbf{B}\right)
-\frac{4\pi~\eta}{c^{2}}~J^{2},
\label{ener_eq}
\end{equation}

\begin{equation}
\frac{\partial{\bf B}}{\partial t} + \nabla\cdot({\bf vB}-{\bf Bv}) = -\nabla\times\left(\frac{4\pi\eta}{c} {\bf J}\right),
\label{ind_eq}
\end{equation}

\begin{equation}
\nabla\cdot{\bf B} = 0.
\label{no_monopoles}
\end{equation}

\noindent Here $\rho$ is the mass density, and ${\bf v}$ is the plasma velocity. The momentum density is ${\bf m}=\rho{\bf v}$; $p$ is the thermal pressure; $p_{t}$ denotes the total (thermal plus magnetic) pressure; ${\bf B}$ is the magnetic field; $I$ is the identity matrix; and the superscript $T$ in equation~(\ref{mom_cons}) denotes the transpose. The total energy density is
\begin{equation}
E_t = \frac{\rho {\bf v}^2}{2} + \rho e + \frac{{\bf B}^2}{8\pi},
\end{equation}
where $e$ is the specific internal energy. The source terms in equations~(\ref{ener_eq}) and (\ref{ind_eq}) include the magnetic diffusivity $\eta$, the current density ${\bf J}=\frac{c}{4\pi}\nabla\times\mathbf{B}$, the speed of light $c$, and the squared current density $J^{2}={\bf J}\cdot{\bf J}$.

To close equations~(\ref{mass_cons})--(\ref{no_monopoles}), we adopt an ideal-gas equation of state, where the internal energy satisfies $\rho e = p/(\gamma - 1)$. The thermal pressure is given by $p = nk_{B}T$, with $n=\rho/(\mu m_p)$ the number density, $k_B$ the Boltzmann constant, $T$ the temperature, and $\mu m_p$ the mean particle mass, where $m_p$ is the proton mass. We set $\mu=1/2$ and $\gamma=5/3$, appropriate for a fully ionized hydrogen plasma. The sound speed is
\begin{equation}
v_s = \sqrt{\frac{\gamma p}{\rho}},
\end{equation}
where $p$ is the proton thermal pressure. The sonic Mach number is $M_s=v/v_s$, and the Alfv\'enic Mach number is $M_a=v/v_A$, where $v$ is the local stellar-wind speed and $v_A$ is the Alfv\'en speed. In all simulations presented here, the stellar wind-magnetosphere interaction occurs in the supersonic and super-Alfv\'enic regimes ($M_s>1$, $M_a>1$).

In this paper, the system of equations~(\ref{mass_cons})--(\ref{no_monopoles}) is solved in a single-fluid, resistive-MHD framework, where non-ideal effects are included through an explicit magnetic diffusivity $\eta$ in the induction and total energy density equations. This formulation permits resistive dissipation and is consistent with a slow-reconnection regime. We note, however, that resistive MHD does not include Hall or kinetic physics; as a result, reconnection rates and the associated energy conversion may differ from those expected when such effects enable fast reconnection.

\subsection{Numerical methods}
\label{sub-section:Methods}

We numerically solve equations~(\ref{mass_cons})--(\ref{no_monopoles}) using the PLUTO code \citep{Mignone_et_al_2007}, which is well suited for modeling high-Mach-number flows in astrophysical plasmas. The equations are integrated in conservative form using a Harten--Lax--van Leer (HLL) Riemann solver and a \texttt{minmod} linear flux limiter. The solenoidal constraint on the magnetic field [equation~(\ref{no_monopoles})] is maintained using the mixed hyperbolic--parabolic divergence-cleaning method of \citet{Dedner_et_al_2002}. The diffusion terms that include diffusivity are evolved separately from the advection terms via operator splitting and are advanced with the super-time-stepping scheme of \citet{Alexiades_et_al_1996}, which efficiently handles the strongly parabolic behavior associated with the high diffusivity values adopted here. In addition, the Courant--Friedrichs--Lewy (CFL) number is set to 0.3, and the parabolic CFL number is fixed at 0.1 in all simulations.

The computational domain is defined on a spherical grid with 128 uniformly spaced cells in the radial direction $r$, 48 cells in the polar direction $\theta\in[0,\pi]$, and 96 cells in the azimuthal direction $\phi\in[0,2\pi]$. The characteristic length scale is set by the planetary radius, $L=R_p=0.918,R_{\oplus}=5.849\times10^{8}$~cm \citep{2016Natur.533..221G}, and the reference stellar-wind speed is varied between $10^{7}$ and $10^{8}$~cm~s$^{-1}$. The radial extent of the domain is a thick spherical shell centered on the exoplanet, with an inner boundary at $R_{\rm in}=2.0~R_p$ and an outer boundary at $R_{\rm out}=30~R_p$. A simplified model of the upper ionosphere occupies the layer between $R_{\rm in}=2.0~R_p$ and $R=2.2~R_p$, resolved with 8 cells. This ionospheric region accounts for field-aligned current effects by prescribing plasma velocities consistent with the electric fields generated by these currents \citep{2018A&A...616A.182V}.

\subsubsection{Initial conditions and boundary conditions}
\label{subsec:Initial_and_boundary_conditions}

The initial and boundary conditions follow the configurations proposed by \citet{2018A&A...616A.182V} and \citet{Wang_et_al_2025}, to which the reader is referred for further details. For completeness, we summarize the key elements here. Between $R_{\rm in}=2.0~R_{p}$ and $R=2.2~R_{p}$, we prescribe plasma inflow driven by field-aligned currents, defining an upper-ionosphere model for TRAPPIST-1e. At the outer boundary, the upstream region is assigned fixed stellar-wind parameters, whereas the downstream region employs zero radial-gradient conditions, $\partial/\partial r=0$, for all plasma variables to minimize artificial numerical reflections associated with electric fields. The initial conditions include a cutoff radius for the interplanetary magnetic field (IMF) at $R_{\textrm{cut}}=6~R_{p}$, where the planetary magnetic pressure exceeds the stellar-wind pressure.

The simulation frame is defined such that the $z$-axis coincides with the planetary magnetic axis and points toward the magnetic north pole. The star lies in the $xy$-plane with $x_{\rm star}>0$, while the $y$-axis completes the right-handed coordinate system.

The characteristic Alfv\'en speed is fixed at $v_{A}=10^{9}$~cm~s$^{-1}$, which sets the simulation time step. The characteristic timescale is $\tau=L/v$, which is of the order of seconds in physical units. All simulations are evolved for $t\approx15$ in code units, corresponding to tens of minutes in physical time.

\subsubsection{Simulation cases}
\label{subsec:Simulation_cases}

We study three cases, hereafter referred to as Case~\#1, Case~\#2, and Case~\#3. In all cases, the stellar magnetic field strength, $B_{\textrm{TRAPPIST-1}}$, is fixed at 600~G, whereas the IMF carried by the stellar wind, $B_{\textrm{SW}}$, and the planetary field strength, $B_{\textrm{TRAPPIST-1e}}$, are varied.

Cases~\#1 and \#2 correspond to super-Alfv\'enic stellar-wind conditions with velocities of $v_{\textrm{SW}}=9.2\times10^{7}$ and $1.0\times10^{8}$~cm~s$^{-1}$, respectively \citep{Pena-Monino_et_al_2024}. In both cases, the stellar-wind magnetic field strength is $2.65\times10^{-3}$~G \citep{Wang_et_al_2025}; the mass densities are $2.51\times10^{-22}$ and $5.16\times10^{-20}$~g~cm$^{-3}$ \citep{2022SpWea..2003164V,Wang_et_al_2025}; and the corresponding temperatures are $2.6\times10^{5}$ and $6.73\times10^{6}$~K \citep{Harbach_et_al_2021}. For Case~\#1, we adopt $T_{\textrm{TRAPPIST-1e}}=1.0\times10^{3}$~K and $B_{\textrm{TRAPPIST-1e}}=0.32$~G, whereas for Case~\#2 we use $T_{\textrm{TRAPPIST-1e}}=800$~K and $B_{\textrm{TRAPPIST-1e}}=0.64$~G \citep{Pena-Monino_et_al_2024,Wang_et_al_2025}.

Case~\#3 mimics an extreme CME-like configuration with $v_{\textrm{SW}}=2.5\times10^{8}$~cm~s$^{-1}$, $B_{\textrm{SW}}=2.65\times10^{-2}$~G, $\rho_{\textrm{SW}}=2.58\times10^{-19}$~g~cm$^{-3}$, and       $T_{\textrm{SW}}=6.73\times10^{6}$~K \citep{Pena-Monino_et_al_2024,Wang_et_al_2025,Harbach_et_al_2021}. For the planet, we set $T_{\textrm{TRAPPIST-1e}}=800$~K and $B_{\textrm{TRAPPIST-1e}}=1.28$~G \citep{Pena-Monino_et_al_2024,Wang_et_al_2025}.

We also compute the dynamic pressure, defined as $p_{\mathrm{dyn}}=\rho v^{2}$ in $\mathrm{dyn~cm^{-2}}$, which provides a convenient measure of the incident momentum flux that drives the wind-planet interaction. We obtain $p_{\mathrm{dyn}}=2.12\times10^{-6}$ for Case~\#1, $p_{\mathrm{dyn}}=5.16\times10^{-4}$ for Case~\#2, and $p_{\mathrm{dyn}}=1.61\times10^{-2}$ for Case~\#3.

The IMF orientation also differs between cases. In Case~\#1, the interplanetary magnetic field (IMF) is aligned with the planetary magnetic axis. In Case~\#2, the IMF is inclined by $15^\circ$ relative to the planetary magnetic axis; accordingly, we adopt $(B_x,B_y,B_z)=(0.2588,0,0.9659)$ in code units, corresponding to a tilt in the $xz$-plane. In Case \#3, the IMF lies in the $xy$-plane, with $(B_x,B_y,B_z)=(-0.707,0.707,0)$.

We adopt an isotropic, spatially constant magnetic diffusivity in equations~(\ref{ener_eq}) and~(\ref{ind_eq}), considering four values: $\eta=0$~cm$^{2}$~s$^{-1}$, $\eta=538.018$~cm$^{2}$~s$^{-1}$, $\eta=5.38018\times10^{8}$~cm$^{2}$~s$^{-1}$, and $\eta=5.38018\times10^{12}$~cm$^{2}$~s$^{-1}$. These values correspond to magnetic Reynolds numbers spanning $R_{\mathrm{m}}\sim10^{14}$ down to $10^{4}$ (specifically, $10^{14}$, $10^{8}$, and $10^{4}$), covering regimes in which advection dominates. In this work, the simulation labeled $\eta=0$~cm$^{2}$~s$^{-1}$ denotes a run with no explicitly prescribed magnetic diffusivity in the induction equation. It should not be interpreted as strictly ideal MHD, because the solution still contains numerical diffusivity that depends on the discretization, Riemann solver, and grid resolution; we therefore use $\eta=0$~cm$^{2}$~s$^{-1}$ as a reference case without explicit diffusivity terms rather than a true zero-diffusion limit.

For our spherical grid ($r\in[2,30]$ resolved by 136 cells (including the 8 cells of the upper ionosphere layer $R_{in}=2.0~R_{p}$ and $R=2.2~R_{p}$), $\theta\in[0,\pi]$ by 48 cells, and $\phi\in[0,2\pi]$ by 96 cells), the characteristic spacings are $\Delta r\simeq(30-2)/136\approx0.206~R_{\rm p}$ and $\Delta\theta=\pi/48\simeq\Delta\phi=2\pi/96\approx0.065$. The corresponding local linear resolutions are $\Delta s_\theta\simeq r~\Delta\theta~R_{\rm p}$ and $\Delta s_\phi\simeq r\sin\theta~\Delta\phi~R_{\rm p}$; for $r\sim3$--$6~R_{\rm p}$ and mid-latitudes this gives a representative $\Delta\sim(1$--$2)\times10^{8}$~$\mathrm{cm}$ (using $R_{\rm p}\approx5.9\times10^{8}$~$\mathrm{cm}$ for TRAPPIST-1e). To estimate the effective (numerical) magnetic diffusivity associated with the discretization, we use standard truncation-error scaling for finite-volume (FV) Godunov-type schemes, where the leading dissipative term scales with a characteristic signal speed times the local mesh spacing \citep[e.g.][]{LeVeque2002}. For the HLL Riemann solver, the dominant numerical dissipation is set by the maximum signal speed, $\lambda_{\max}\sim |v_n|+c_f$, motivating an effective diffusivity of order
$\eta_{\rm num}\sim C~\lambda_{\max}~\Delta$, where $C$ depends on the reconstruction and slope limiter \citep{Harten1983,Sweby1984}. Given our use of the \texttt{minmod} limiter, we adopt $C\simeq0.3$, consistent with limiter-dependent numerical-diffusion estimates \citep[e.g.][]{JCP_TVD_NumDiff_2013}. For typical magnetospheric values $(|v_n|+c_f)\sim10^{7}$--$10^{8}$~$\mathrm{cm~s^{-1}}$, this yields $\eta_{\rm num}\sim10^{15}$--$10^{16}$~$\mathrm{cm^{2}~s^{-1}}$ in the regions of strongest gradients (with expected spatial variability).

For close-in M-dwarf exoplanets, strong stellar magnetic fields, frequent flaring activity, and high stellar-wind densities are expected to enhance reconnection and the effective magnetic diffusivity relative to Solar System conditions \citep[e.g.][]{Cohen_2014,Garraffo_2017}. We therefore consider three simulation cases spanning different upstream forcings and four values of $\eta$. In all simulations, the reference case for comparison is $\eta=0$. Table~\ref{tab:parameters} summarizes the parameters adopted for the three cases analyzed in this paper.

\begin{table}
	\centering
	\caption{Input parameters used for the three simulation cases}
	\label{tab:parameters}
	\begin{tabular}{lccr} 
		\hline
		Parameter & Case \#1 & Case \#2 & Case \#3\\
		\hline
		$v_{\textrm{SW}}$ (cm s$^{-1}$) & $9.2\times10^{7}$ & $1\times10^{8}$ & $2.5\times10^{8}$ \\
        $B_{\textrm{SW}}$ (G) & $2.65\times10^{-3}$  & $2.65\times10^{-3}$ & $2.65\times10^{-2}$\\
		$\rho_{\textrm{SW}}$ (g cm$^{-3}$) & $2.51\times10^{-22}$ & $5.16\times10^{-20}$ & $2.58\times10^{-19}$\\
        $T_{\textrm{SW}}$ (K) & $2.6\times10^{5}$ & $6.73\times10^{6}$ & $6.73\times10^{6}$ \\
        $p_{\textrm{dyn}}$ (dyn cm$^{-2}$) & $2.12\times10^{-6}$ & $5.16\times10^{-4}$
        & $1.61\times10^{-2}$  \\
        $T_{\textrm{TRAPPIST-1e}}$ (K) & $1\times10^{3}$ & 800 & 800 \\
        $B_{\textrm{TRAPPIST-1e}}$ (G) & 0.32 & 0.64 & 1.28 \\
		\hline
	\end{tabular}
\end{table}

\section{Results of the numerical simulations}
\label{sec:Results_numerical_simulations}

In this section, we present representative results for the three simulation cases at magnetic diffusivity values $\eta=0$, and $5.38018\times10^{12}$~cm$^{2}$~s$^{-1}$. We focus on equatorial and meridional maps of mass density, dynamic pressure, total energy density, total Poynting flux, and the divergence of the total Poynting flux, which together characterize the system's global structure. In all figures, the white circle marks the inner boundary of the computational domain ($R_{\rm in}=2~R_p$) and does not necessarily correspond to the exoplanetary body.

\subsection{Case \#1}
\label{sub_sec:Results_case1}

\begin{figure*}
\centering
\textbf{Case \#1}
\begin{center}
\begin{subfigure}[b]{0.3\textwidth}
\centering
\caption{}
\includegraphics[width=\columnwidth]{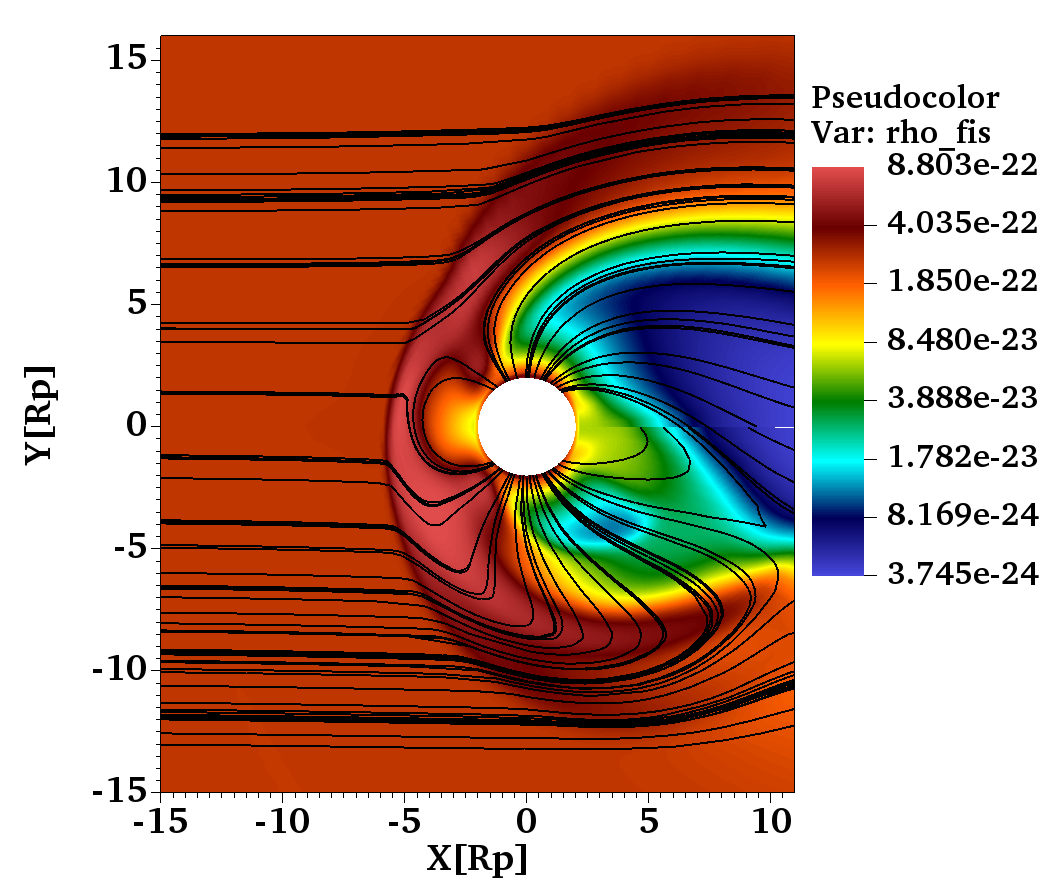}
\end{subfigure}
\begin{subfigure}[b]{0.3\textwidth}
\centering
\caption{}
\includegraphics[width=\columnwidth]{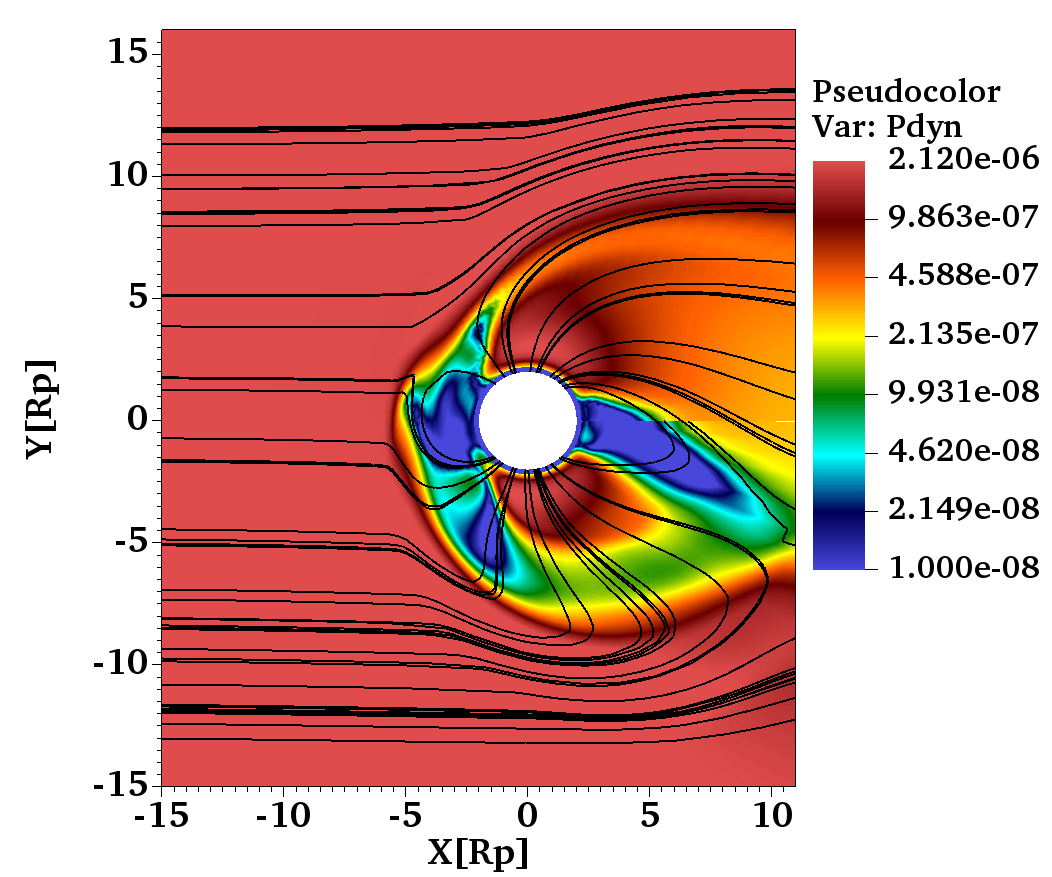}
\end{subfigure}
\begin{subfigure}[b]{0.3\textwidth}
\centering
\caption{}
\includegraphics[width=\columnwidth]{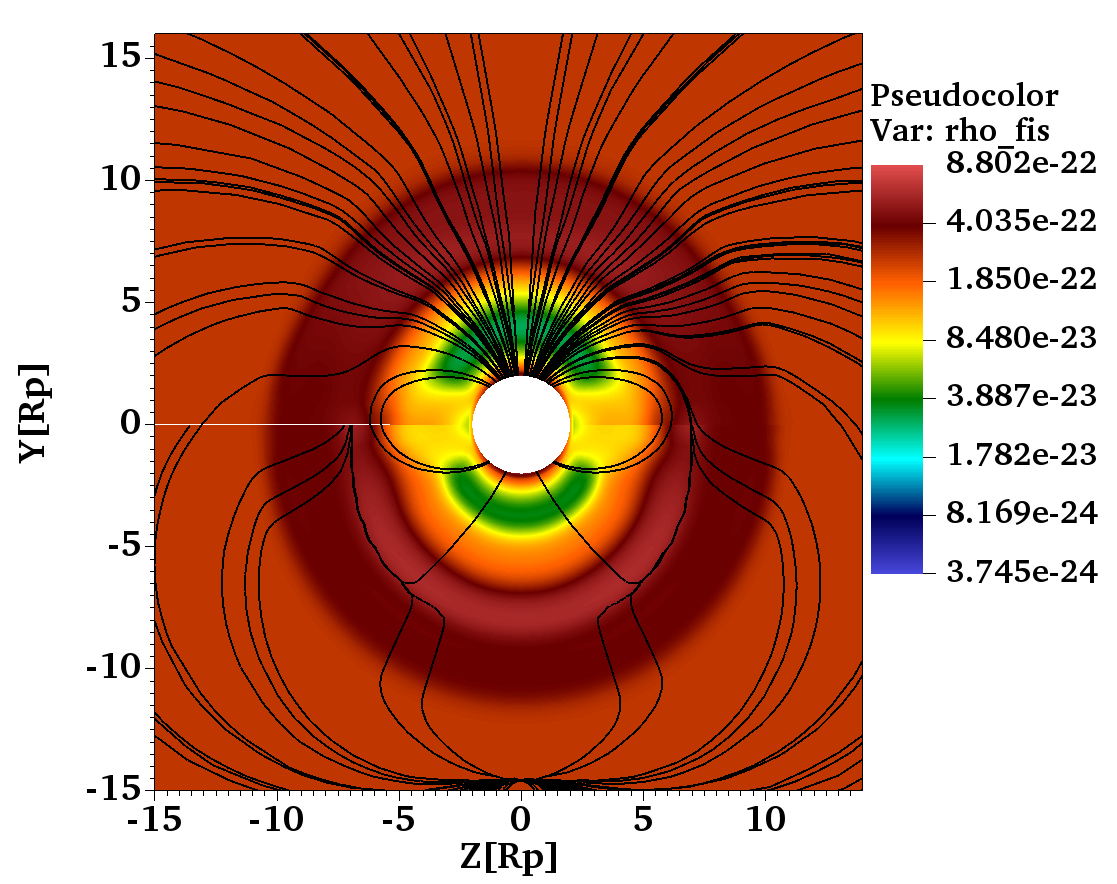}
\end{subfigure} 
\begin{subfigure}[b]{0.3\textwidth}
\centering
\caption{}
\includegraphics[width=\columnwidth]{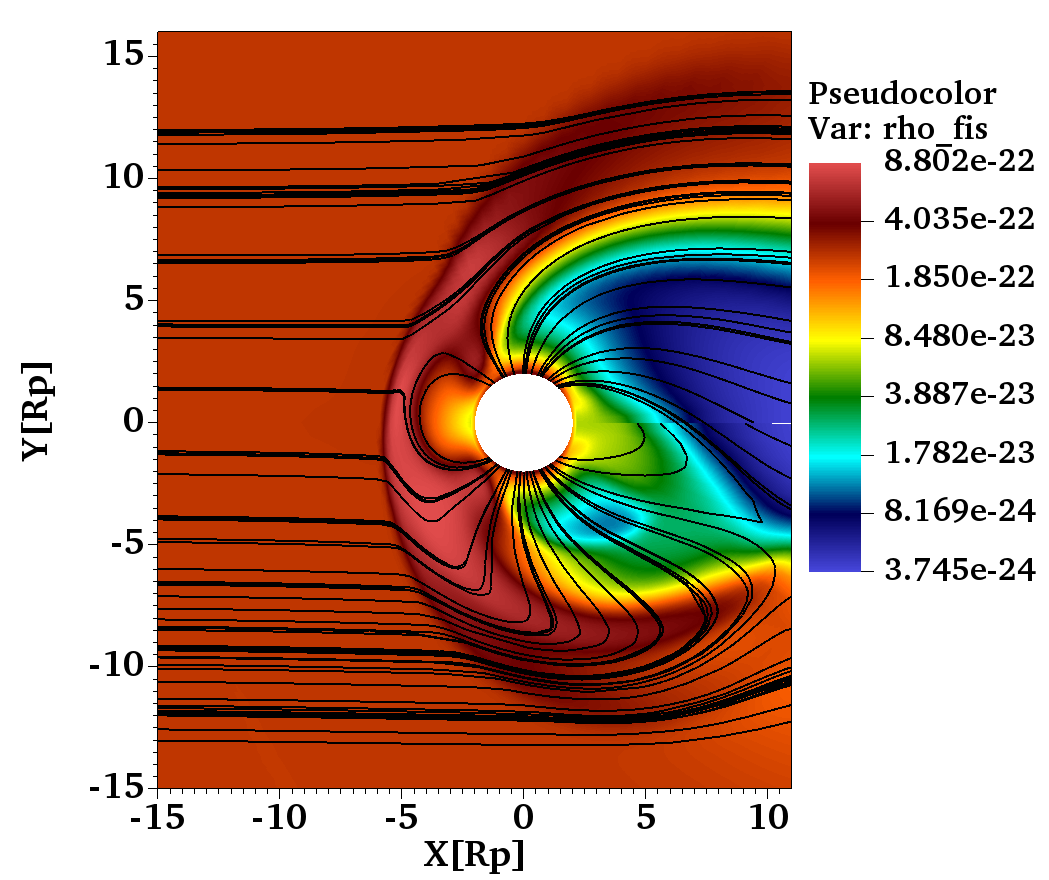}
\end{subfigure}
\begin{subfigure}[b]{0.3\textwidth}
\centering
\caption{}
\includegraphics[width=\columnwidth]{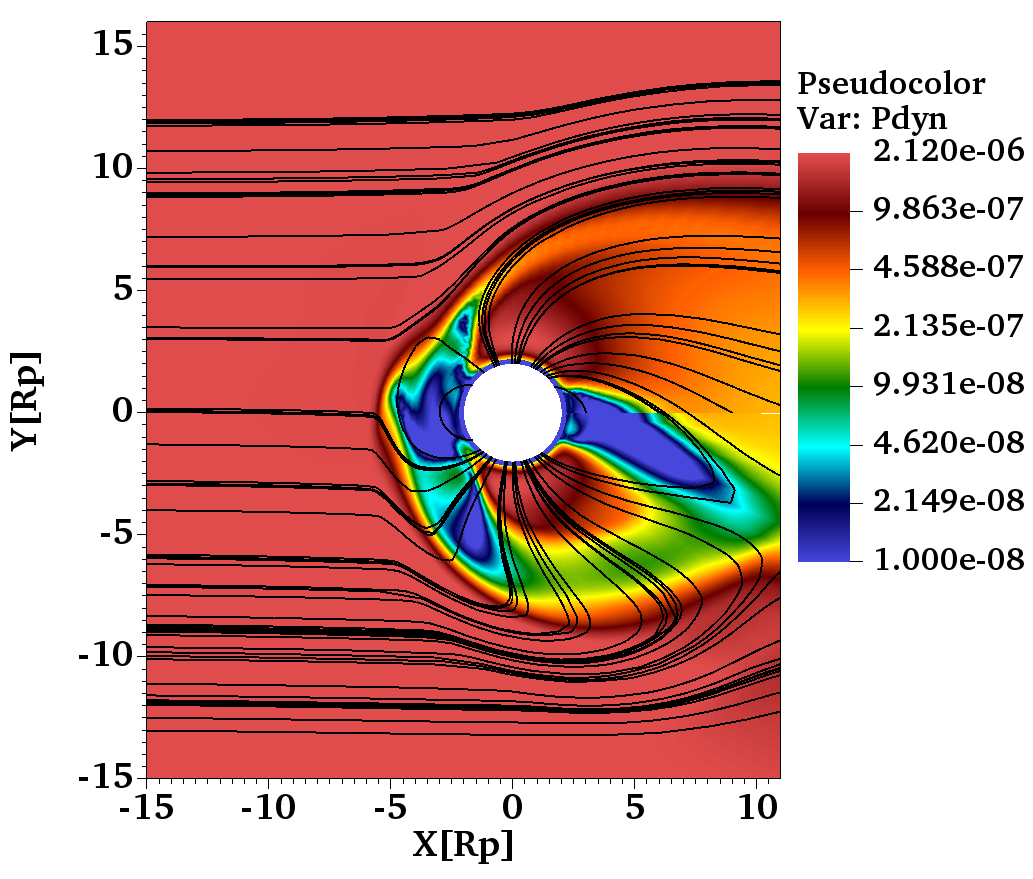}
\end{subfigure}
\begin{subfigure}[b]{0.3\textwidth}
\centering
\caption{}
\includegraphics[width=\columnwidth]{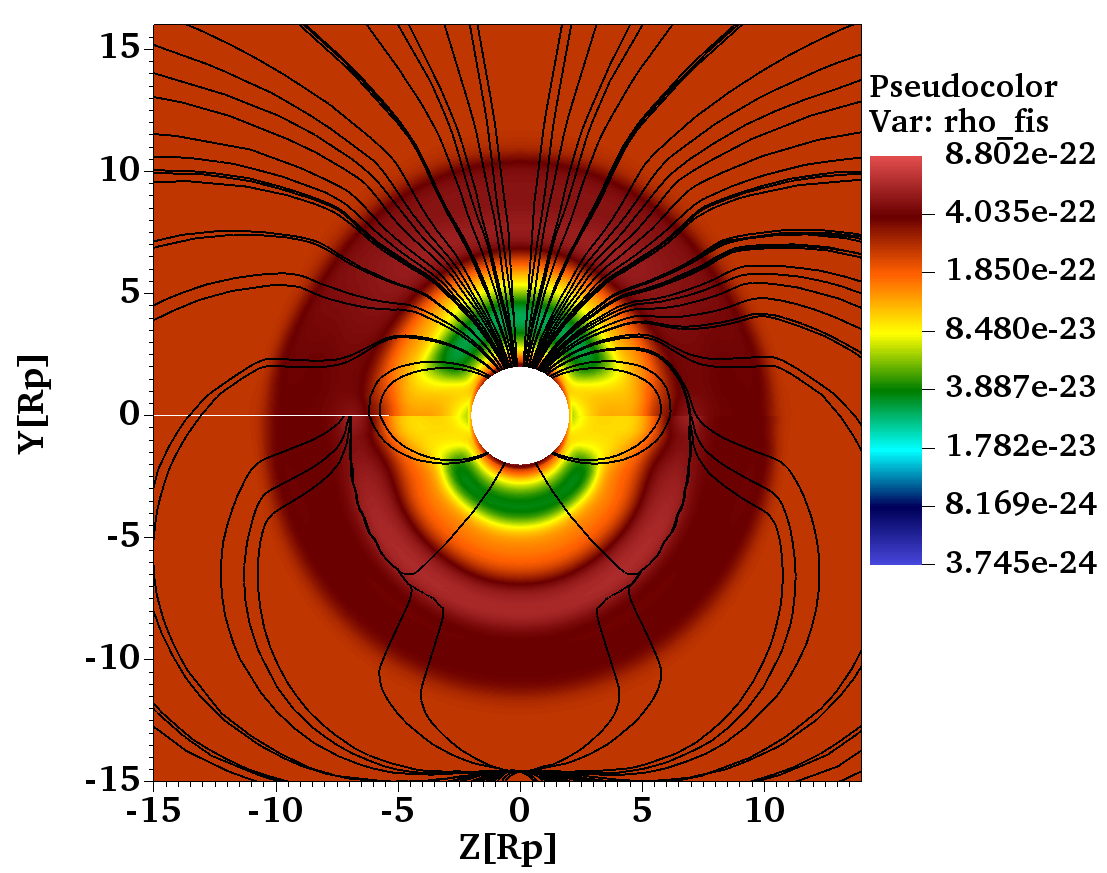}
\end{subfigure}
\caption{Case~\#1. Equatorial-plane maps of mass density $\rho$ (panels a and d; g cm$^{-3}$) and dynamic pressure $p_{\rm dyn}$ (panels b and e; dyn cm$^{-2}$), together with meridional-plane maps of $\rho$ (panels c and f). Black curves show the magnetic field lines. The top row corresponds to $\eta=0$ cm$^{2}$ s$^{-1}$, and the bottom row to $\eta=5.38018\times10^{12}$ cm$^{2}$ s$^{-1}$.}
\label{fig:Results_case1}
\end{center}
\end{figure*}

Figure~\ref{fig:Results_case1} presents the Case~\#1 stellar wind--planet interaction for $\eta = 0$~cm$^{2}$~s$^{-1}$ (panels a--c) and $\eta = 5.38018 \times 10^{12}$~cm$^{2}$~s$^{-1}$ (panels d--f). Panels a, b, d, and e show equatorial maps of the mass density, $\rho$ (g~cm$^{-3}$), and dynamic pressure, $p_{\rm dyn}$ (dyn~cm$^{-2}$), while panels c and f show meridional maps of $\rho$; in all panels, black curves indicate the overplotted magnetic field lines. In both cases, the stellar wind compresses the planetary obstacle and forms a well-defined magnetospheric cavity, visible as a low-$\rho$, low-$p_{\rm dyn}$ region surrounding the planet. The magnetic field lines are compressed upstream, drape around the magnetospheric boundary, and remain closed near the planet, thereby outlining the dipole-dominated inner region. As the magnetic diffusivity increases, the explicit resistive term smooths the sharp magnetic gradients at the wind--magnetosphere interface, broadening the current sheets and making the regions of very low magnetic-field strength more spatially extended. Although the overall morphology is similar in both cases, the finite-diffusivity solution is slightly smoother, with broader gradients in $\rho$ and $p_{\rm dyn}$ and a more diffuse field-line draping pattern, whereas the $\eta = 0$ case preserves sharper transitions and more structured plasma features.

\subsection{Case \#2}
\label{sub_sec:Results_case2}

\begin{figure*}
\centering
\textbf{Case \#2}
\begin{center}
\begin{subfigure}[b]{0.3\textwidth}
\centering
\caption{}
\includegraphics[width=\columnwidth]{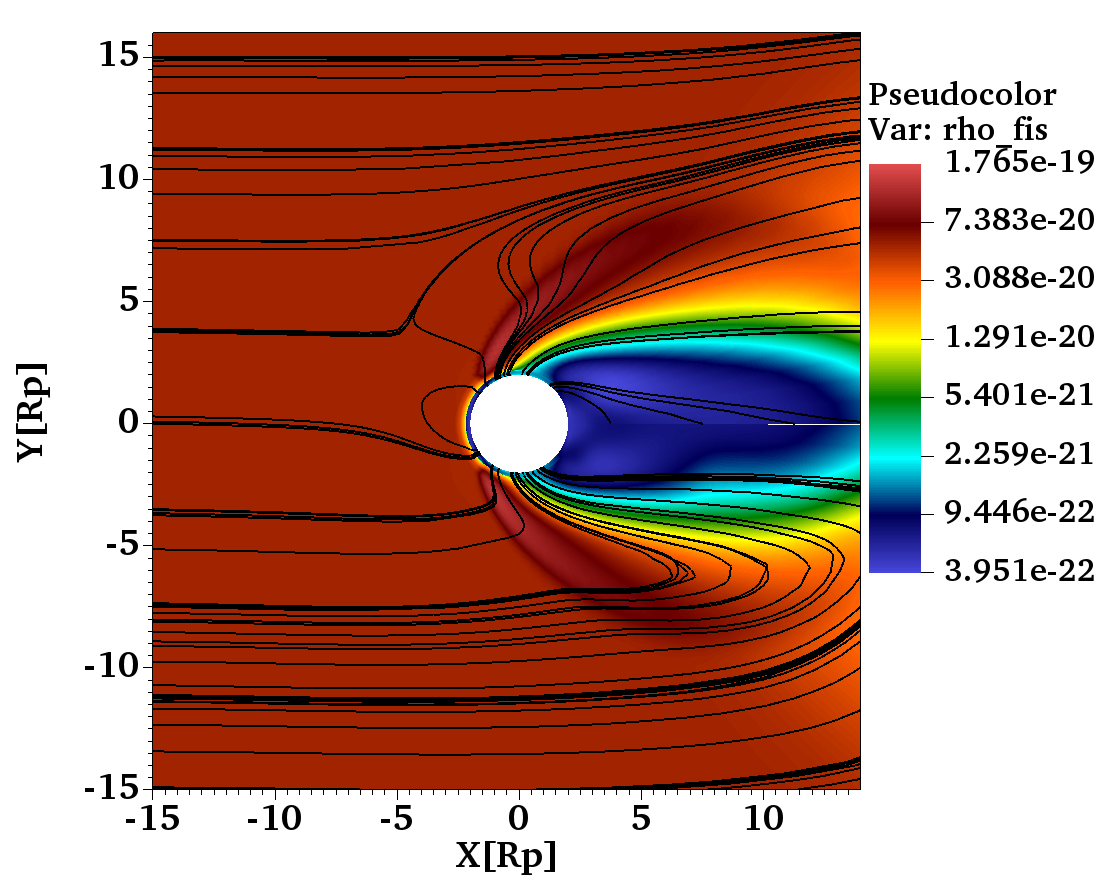}
\end{subfigure}
\begin{subfigure}[b]{0.3\textwidth}
\centering
\caption{}
\includegraphics[width=\columnwidth]{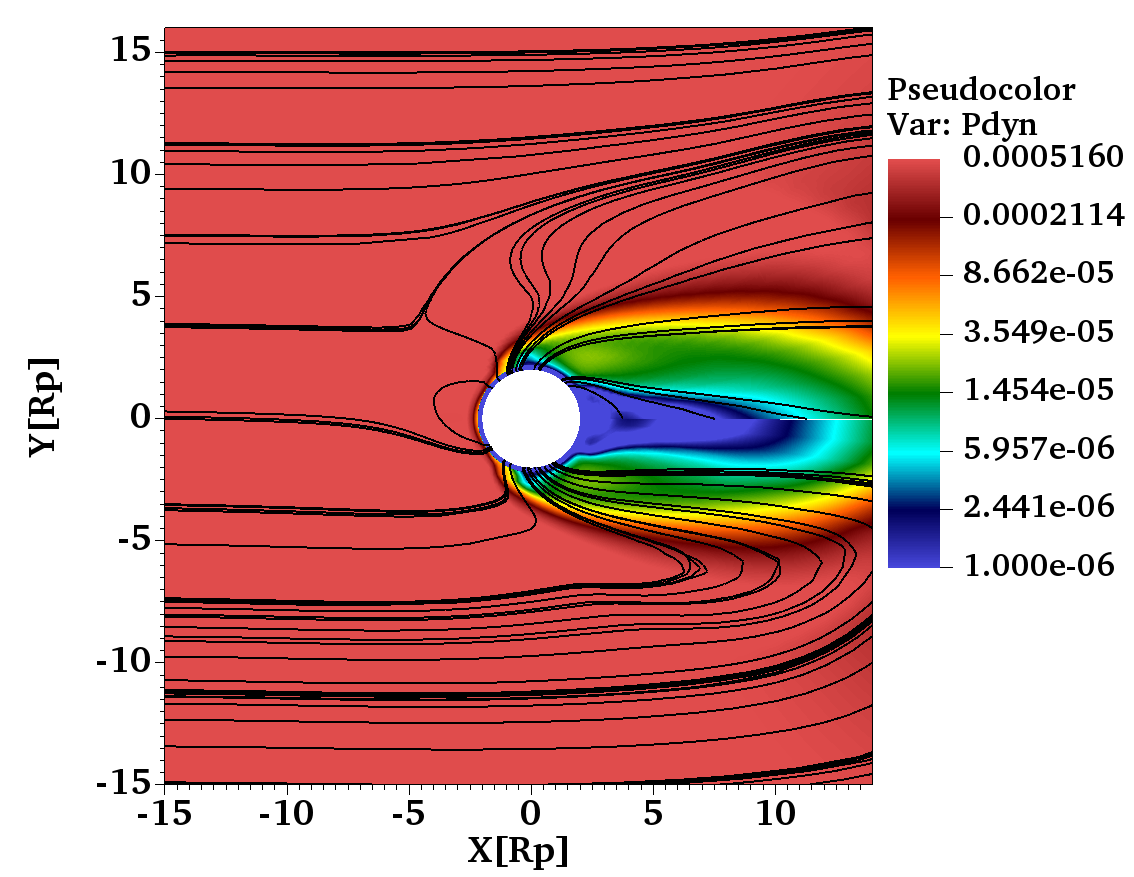}
\end{subfigure}
\begin{subfigure}[b]{0.3\textwidth}
\centering
\caption{}
\includegraphics[width=\columnwidth]{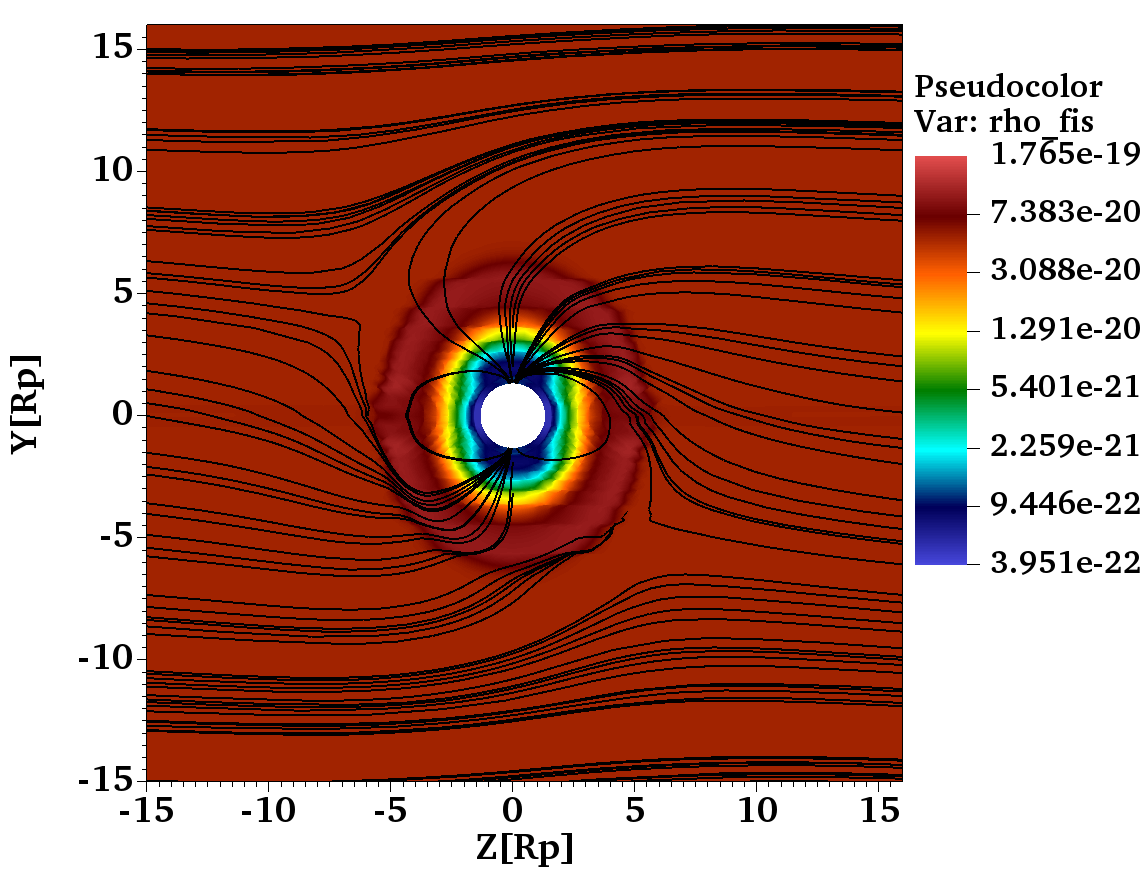}
\end{subfigure}
\begin{subfigure}[b]{0.3\textwidth}
\centering
\caption{}
\includegraphics[width=\columnwidth]{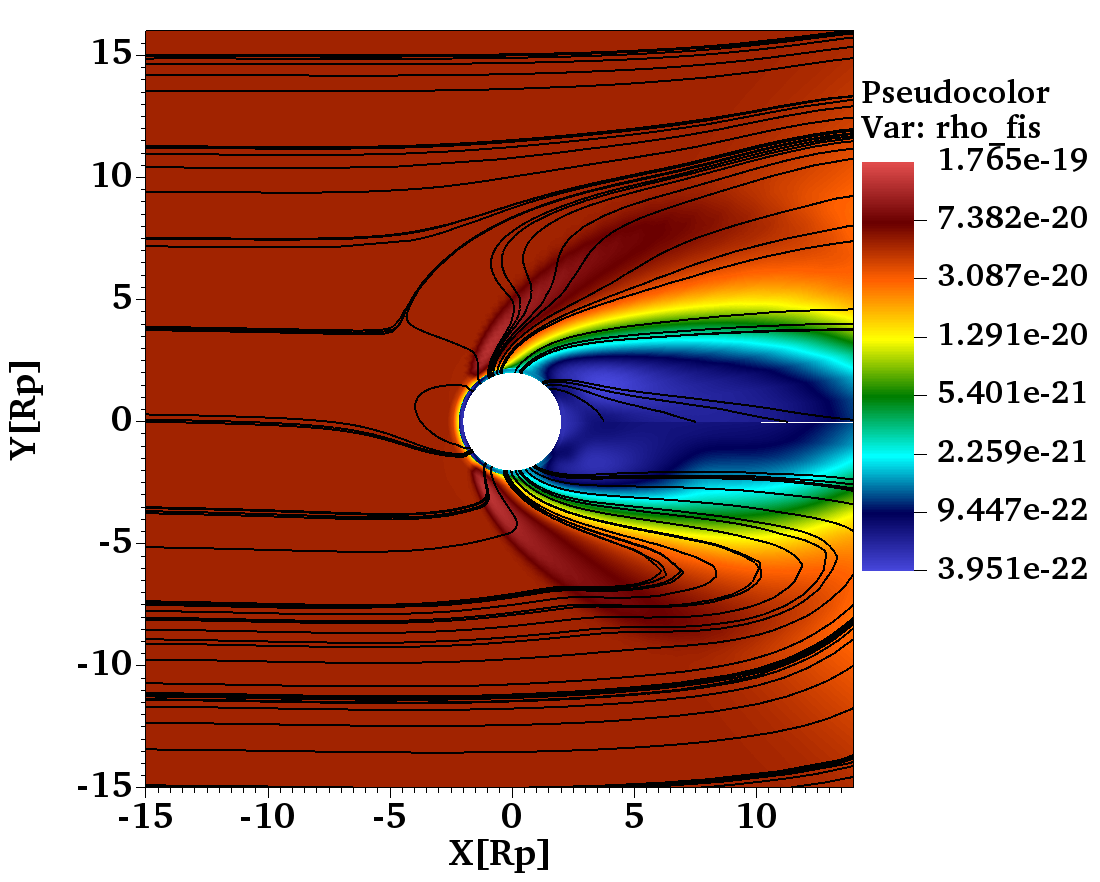}
\end{subfigure}
\begin{subfigure}[b]{0.3\textwidth}
\centering
\caption{}
\includegraphics[width=\columnwidth]{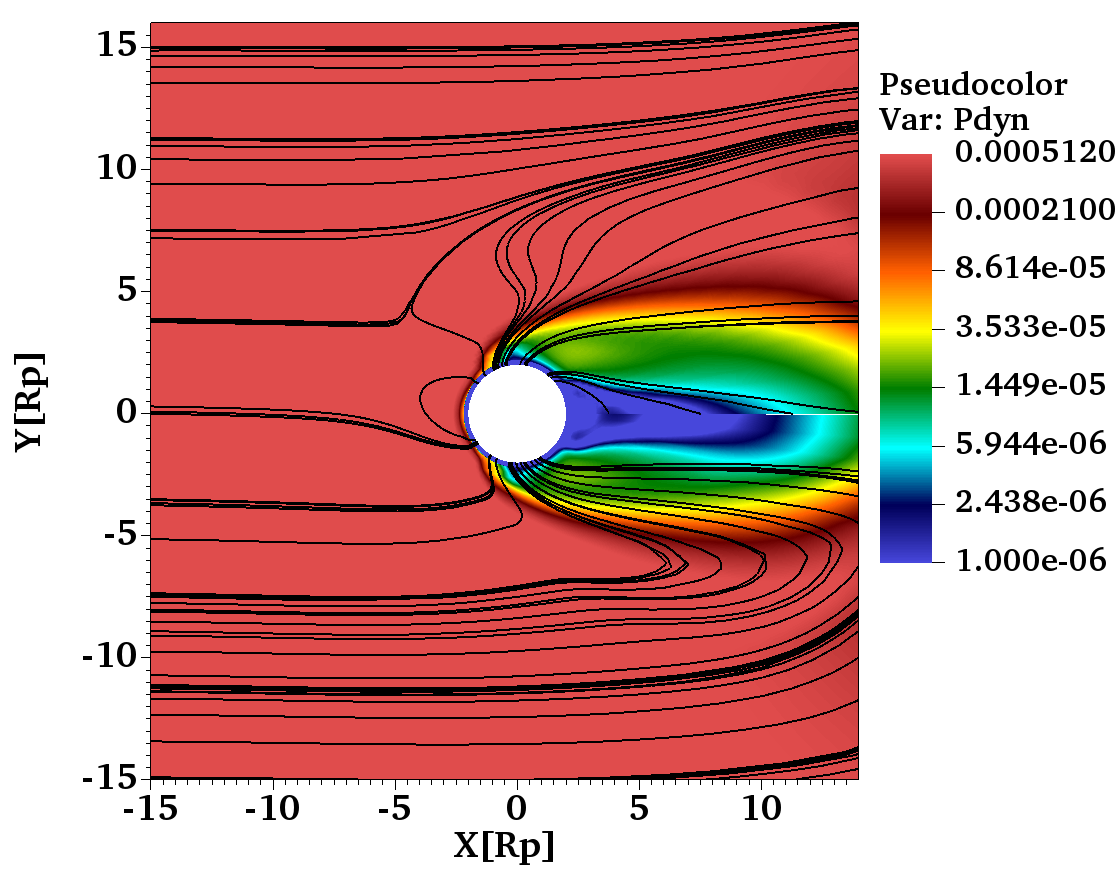}
\end{subfigure}
\begin{subfigure}[b]{0.3\textwidth}
\centering
\caption{}
\includegraphics[width=\columnwidth]{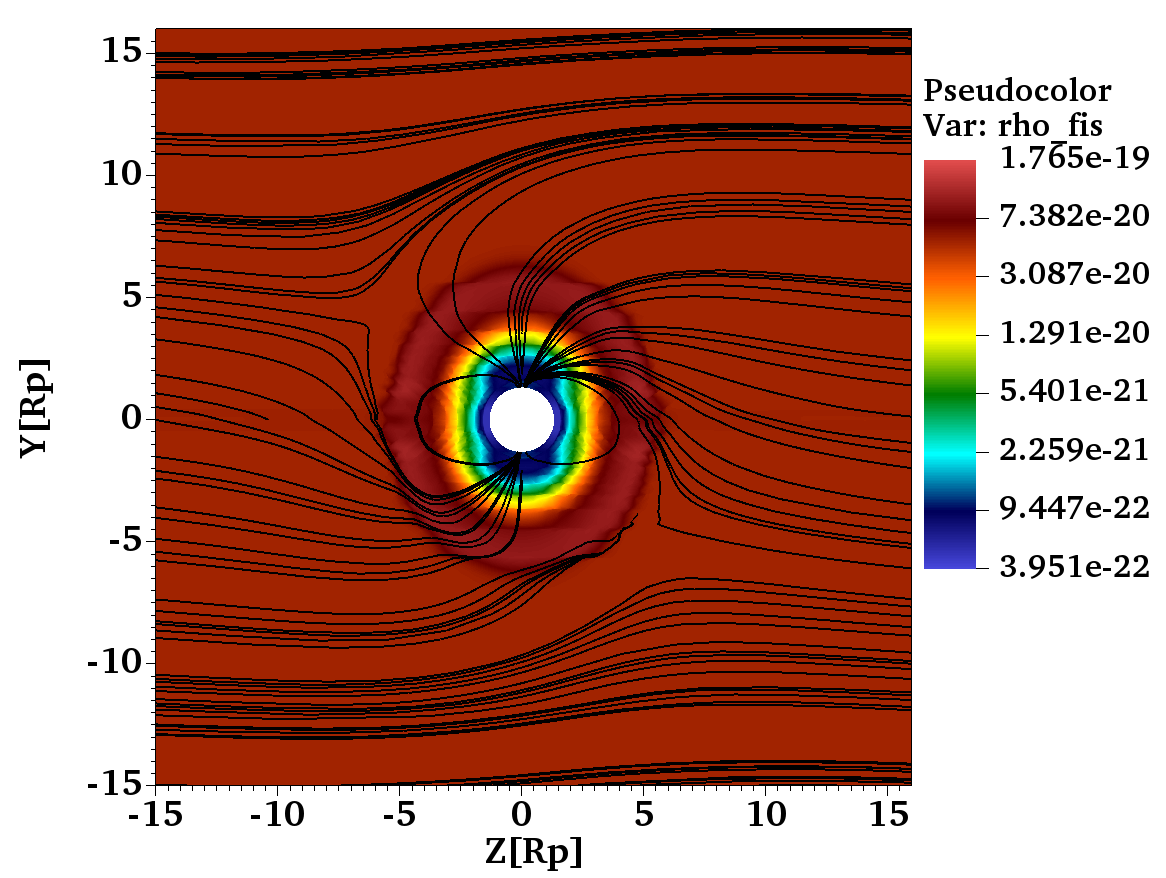}
\end{subfigure}
\caption{Case~\#2. Equatorial-plane maps of mass density $\rho$ (panels a and d; g cm$^{-3}$) and dynamic pressure $p_{\rm dyn}$ (panels b and e; dyn cm$^{-2}$), with magnetic field lines overplotted in black. Panels c and f show meridional-plane maps of $\rho$ at $x=1.5\,R_{\rm p}$. The top row corresponds to $\eta=0$ cm$^{2}$ s$^{-1}$, and the bottom row to $\eta=5.38018\times10^{12}$ cm$^{2}$ s$^{-1}$.}
\label{fig:Results_case2}
\end{center}
\end{figure*}

Figure~\ref{fig:Results_case2} depicts a more strongly confined interaction in Case~\#2, with $\eta = 0$~cm$^{2}$~s$^{-1}$ in panels a--c and $\eta = 5.38018 \times 10^{12}$~cm$^{2}$~s$^{-1}$ in panels d--f. In the equatorial plane, the stellar wind squeezes the planetary field into a compact dayside obstacle and drives a long, narrow downstream cavity, seen in both $\rho$ and $p_{\rm dyn}$ as a sharply depleted channel extending along the tail. The magnetic field lines bend tightly around the flanks of the planet and are swept into a stretched, tail-aligned configuration, indicating strong magnetic confinement and tail formation. The meridional cuts at $x = 1.5~R_{\rm p}$ show that this cavity remains compact in cross-section and is surrounded by compressed plasma. As $\eta$ increases, the explicit resistive term smooths the sharp magnetic gradients at the wind--magnetosphere interface, causing the current sheets to broaden and the regions of very low magnetic-field strength to become more spatially extended. The higher-$\eta$ solution therefore appears slightly smoother, with softer cavity boundaries and less abrupt field-line curvature than the $\eta = 0$ case.

\subsection{Case \#3}
\label{sub_sec:Results_case3}

\begin{figure*}
\centering
\textbf{Case \#3}
\begin{center}
\begin{subfigure}[b]{0.3\textwidth}
\centering
\caption{}
\includegraphics[width=\columnwidth]{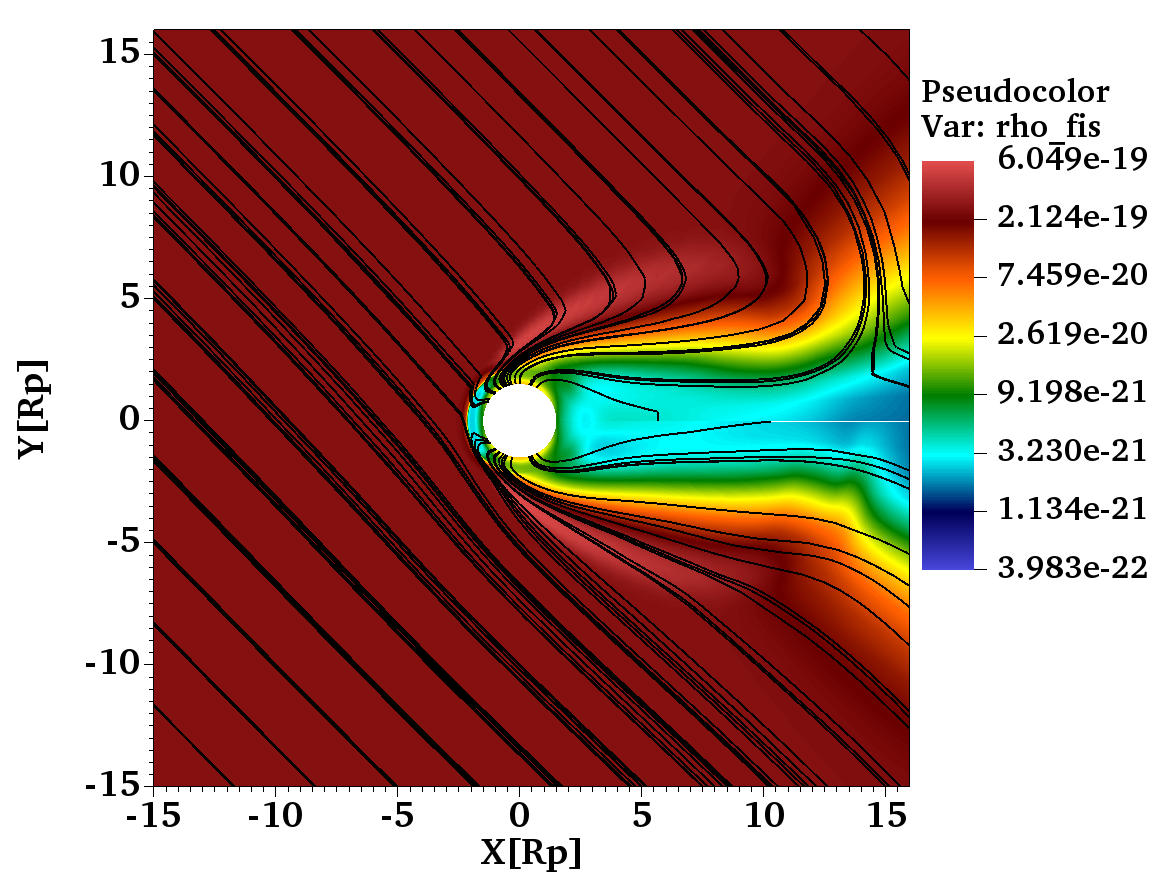}
\end{subfigure}
\begin{subfigure}[b]{0.3\textwidth}
\centering
\caption{}
\includegraphics[width=\columnwidth]{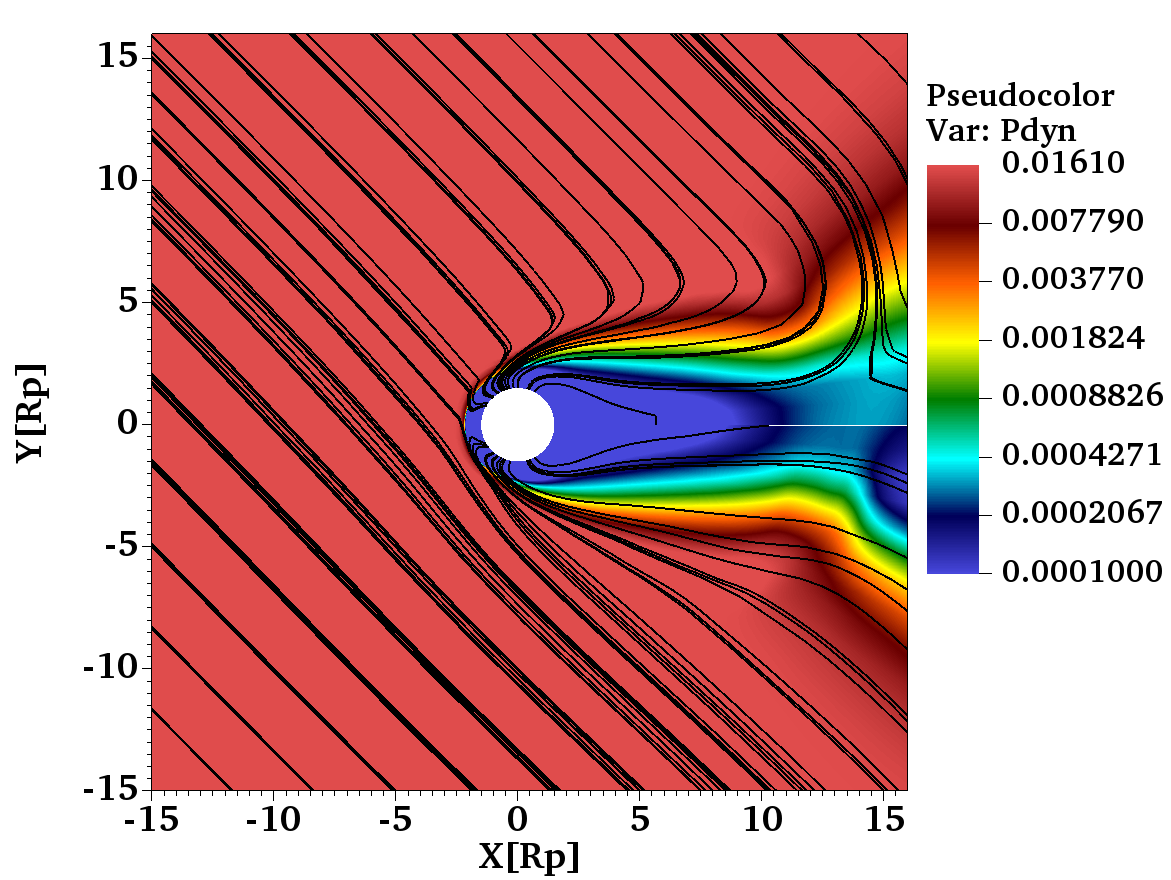}
\end{subfigure}
\begin{subfigure}[b]{0.3\textwidth}
\centering
\caption{}
\includegraphics[width=\columnwidth]{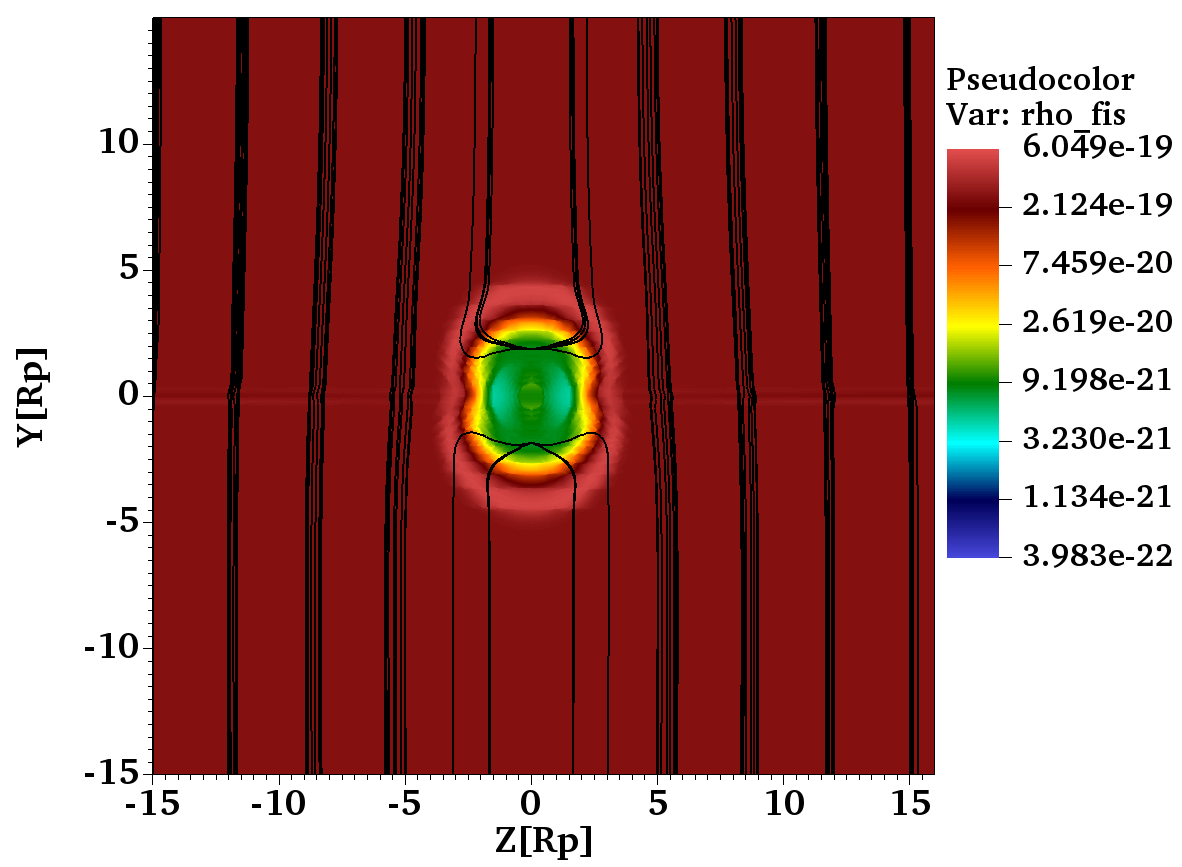}
\end{subfigure}
\begin{subfigure}[b]{0.3\textwidth}
\centering
\caption{}
\includegraphics[width=\columnwidth]{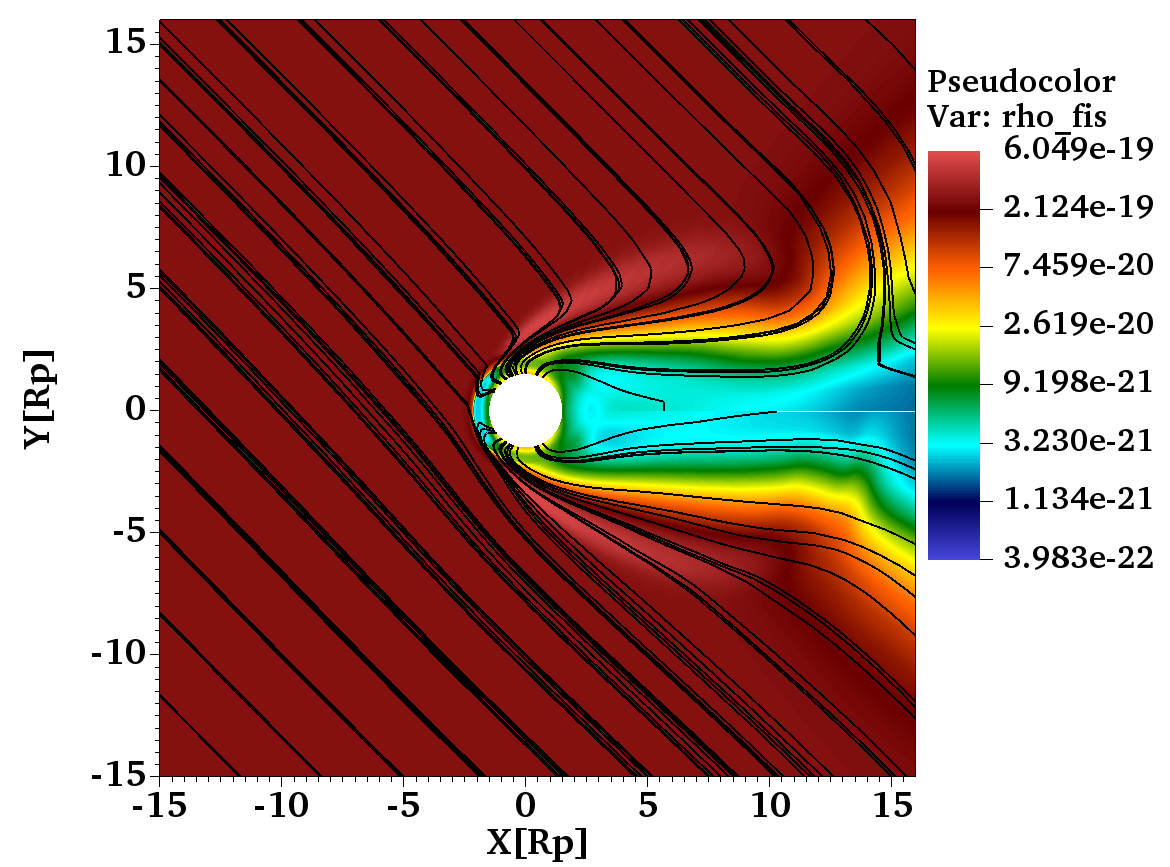}
\end{subfigure}
\begin{subfigure}[b]{0.3\textwidth}
\centering
\caption{}
\includegraphics[width=\columnwidth]{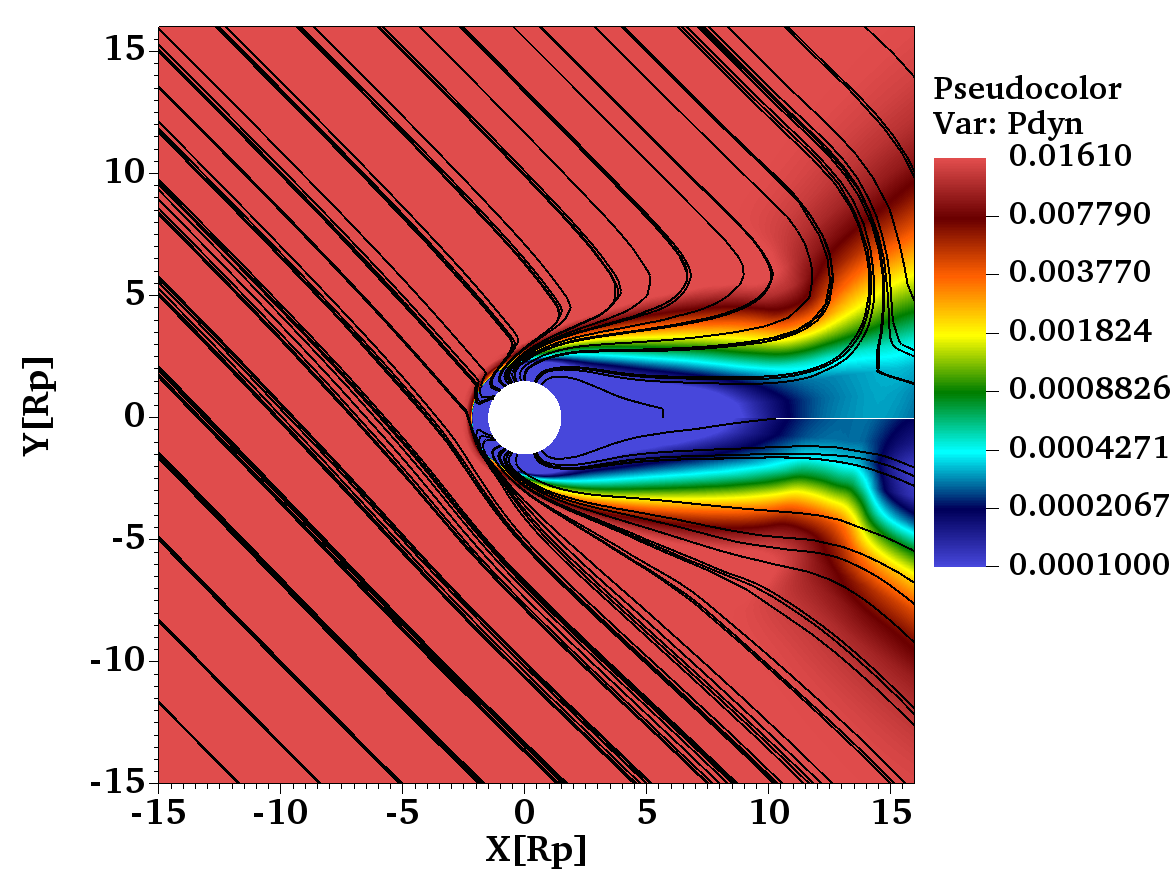}
\end{subfigure}
\begin{subfigure}[b]{0.3\textwidth}
\centering
\caption{}
\includegraphics[width=\columnwidth]{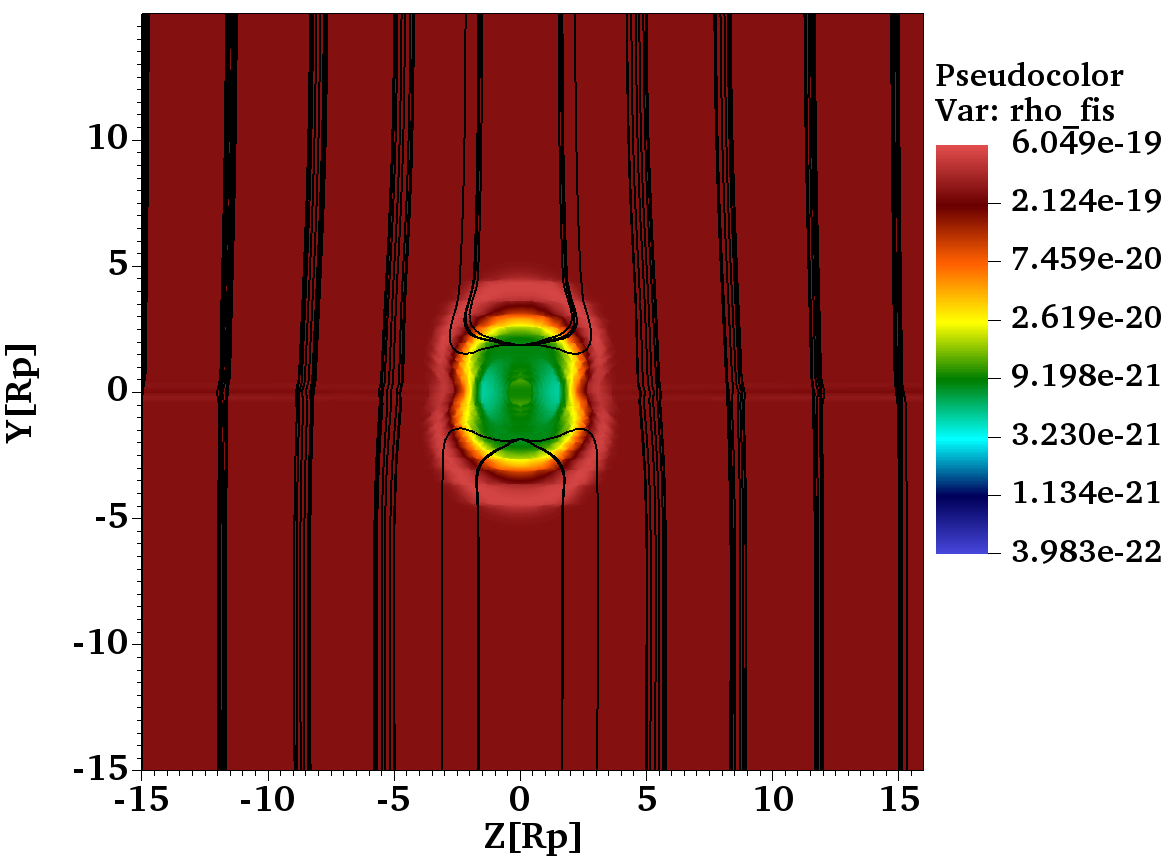}
\end{subfigure}
\caption{Case~\#3. Equatorial-plane maps of mass density $\rho$ (panels a and d; g cm$^{-3}$) and dynamic pressure $p_{\rm dyn}$ (panels b and e; dyn cm$^{-2}$), with magnetic field lines overplotted in black. Panels c and f show meridional-plane maps of $\rho$ at $x=1.5~R_{\rm p}$. The top row corresponds to $\eta=0$ cm$^{2}$ s$^{-1}$, and the bottom row to $\eta=5.38018\times10^{12}$ cm$^{2}$ s$^{-1}$.}
\label{fig:Results_case3}
\end{center}
\end{figure*}

Figure~\ref{fig:Results_case3} displays the Case~\#3 interaction for $\eta=0$~cm$^{2}$~s$^{-1}$ (panels a--c) and $\eta=5.38018\times10^{12}$~cm$^{2}$~s$^{-1}$ (panels d--f). Unlike Cases~\#1 and \#2, the flow here drives a markedly more collimated response, with $\rho$ and $p_{\rm dyn}$ revealing a deep, narrow depleted channel that extends far downstream of the planet. The magnetic field lines are strongly swept back and remain tightly bundled along the flanks of this cavity, indicating substantial field-line draping and efficient confinement of the wake. The meridional cuts at $x=1.5\,R_{\rm p}$ show a compact low-density core bounded by a thin compressed shell, consistent with a strongly compressed obstacle under intense wind forcing. Differences between the two diffusivities are minimal at the scale shown, although the $\eta=5.38018\times10^{12}$~cm$^{2}$~s$^{-1}$ solution exhibits slightly smoother cavity edges and weaker field-line crowding than the $\eta=0$ case.

\subsection{Estimation of radio emission}
\label{sub_sec:emission}

To estimate the radio emission generated by the bow shock and magnetopause in the interaction between the stellar wind and the exoplanetary magnetosphere, we follow the approach of \citet{Pena-Monino_et_al_2024} and define the emitted radio power as
\begin{equation}
P_{R} = \beta~P_{B},
\label{eq:PR_def}
\end{equation}
where $\beta$ is an empirical parameter equal to $2\times10^{-3}$ \citep{Zarka2018}, and $P_{B}$ is a proxy derived from the divergence of the Poynting flux over a prescribed dayside interaction volume $V$ (the bow-shock layer, the magnetopause coupling layer, or their union). In our notation,
\begin{equation}
P_{B} \equiv \int_{V}\nabla\cdot\mathbf{S}_{\rm ideal}~dV,
\qquad
\mathbf{S}_{\rm ideal} \equiv -\frac{(\mathbf{v}\times\mathbf{B})\times\mathbf{B}}{4\pi}.
\label{eq:PB_ideal}
\end{equation}
By Gauss' theorem, $P_{B}=\oint_{\partial V}\mathbf{S}\cdot d\mathbf{A}$, so the sign of $P_{B}$ depends on the adopted orientation of the outward normal on $\partial V$. With the standard outward-normal convention, $P_{B}>0$ corresponds to a net outward Poynting flux through $\partial V$, whereas $P_{B}<0$ corresponds to a net inward Poynting flux (net electromagnetic energy entering $V$). In this paper, we consider a net outward flux alongside the spatial distribution of $\nabla\cdot\mathbf{S}$ maps.

In resistive MHD, the electric field in cgs units is given by the generalized Ohm's law,
\begin{equation}
\mathbf{E} = -\frac{\mathbf{v}\times\mathbf{B}}{c} + \frac{\eta}{c}\mathbf{J},
\qquad
\mathbf{J}=\frac{c}{4\pi}\nabla\times\mathbf{B},
\label{eq:ohm_law_resistive}
\end{equation}
so that the total Poynting flux $\mathbf{S}_{\rm total}= \frac{c}{4\pi}\mathbf{E}\times\mathbf{B}$ can be decomposed as
\begin{equation}
\mathbf{S}_{\rm total}=\mathbf{S}_{\rm ideal}+\mathbf{S}_{\rm res},
\qquad
\mathbf{S}_{\rm res}=\frac{\eta}{4\pi}(\nabla\times\mathbf{B})\times\mathbf{B}.
\label{eq:S_total_decomp}
\end{equation}
We therefore generalize equation~(\ref{eq:PB_ideal}) as
\begin{equation}
P_{B} \equiv \int_{V}\nabla\cdot\mathbf{S}_{\rm total}~dV
= \int_{V}\nabla\cdot\left(\mathbf{S}_{\rm ideal}+\mathbf{S}_{\rm res}\right)~dV.
\label{eq:PB_total}
\end{equation}

We stress that $\nabla\cdot\mathbf{S}_{\rm total}$ is, in general, a flux-divergence term in the electromagnetic energy equation; that is, it can take either sign and may exhibit alternating-sign structures within the coupling layer. Consequently, $P_{B}$ should be interpreted as a signed net Poynting-flux budget over $V$, whose value can reflect both energy conversion and redistribution across $\partial V$.

\begin{figure*}
\centering
\begin{subfigure}[b]{0.4\textwidth}
\centering
\caption{}
\includegraphics[width=\columnwidth]{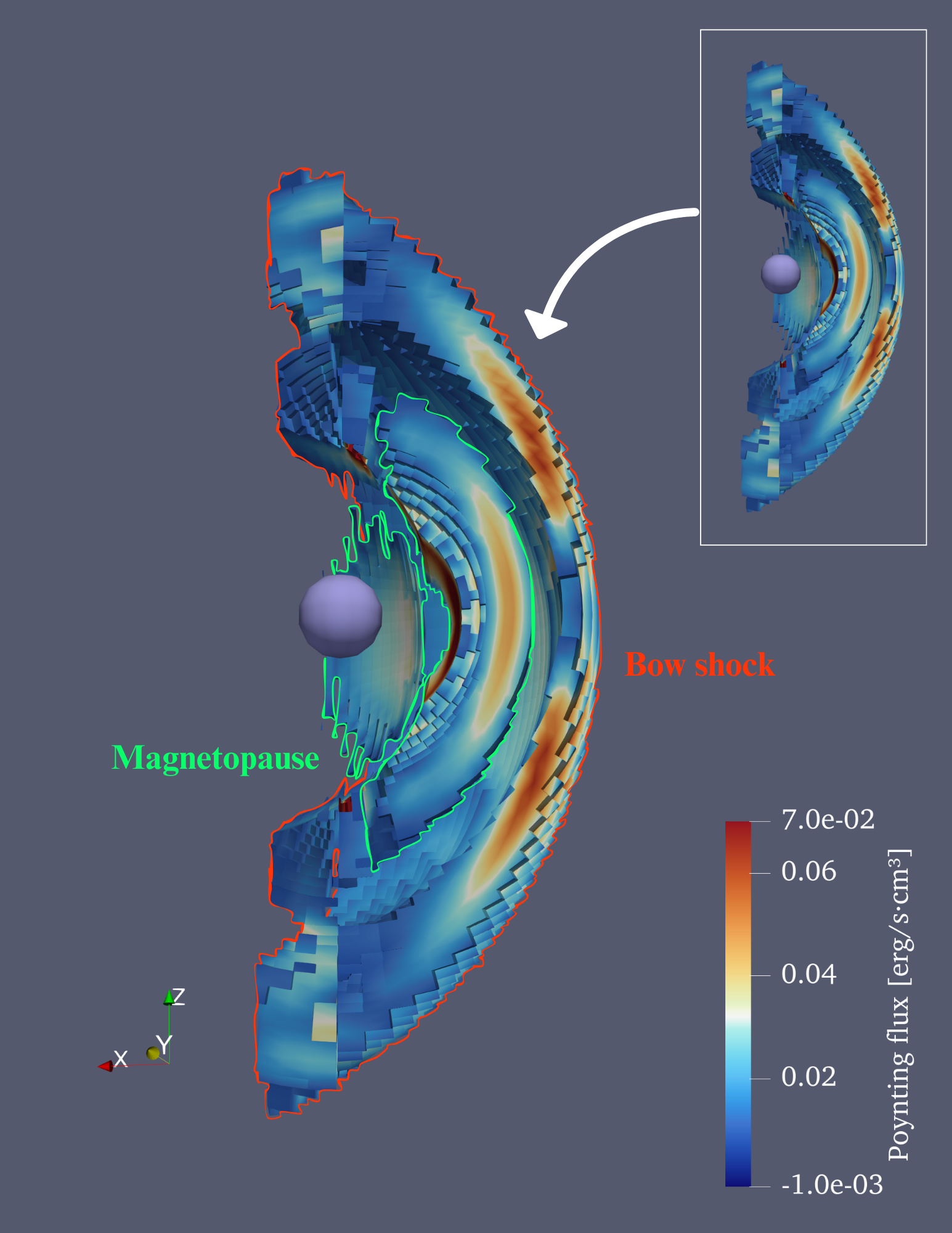}
\end{subfigure}
\hspace{0.1cm}
\begin{subfigure}[b]{0.4\textwidth}
\centering
\caption{}
\includegraphics[width=\columnwidth]{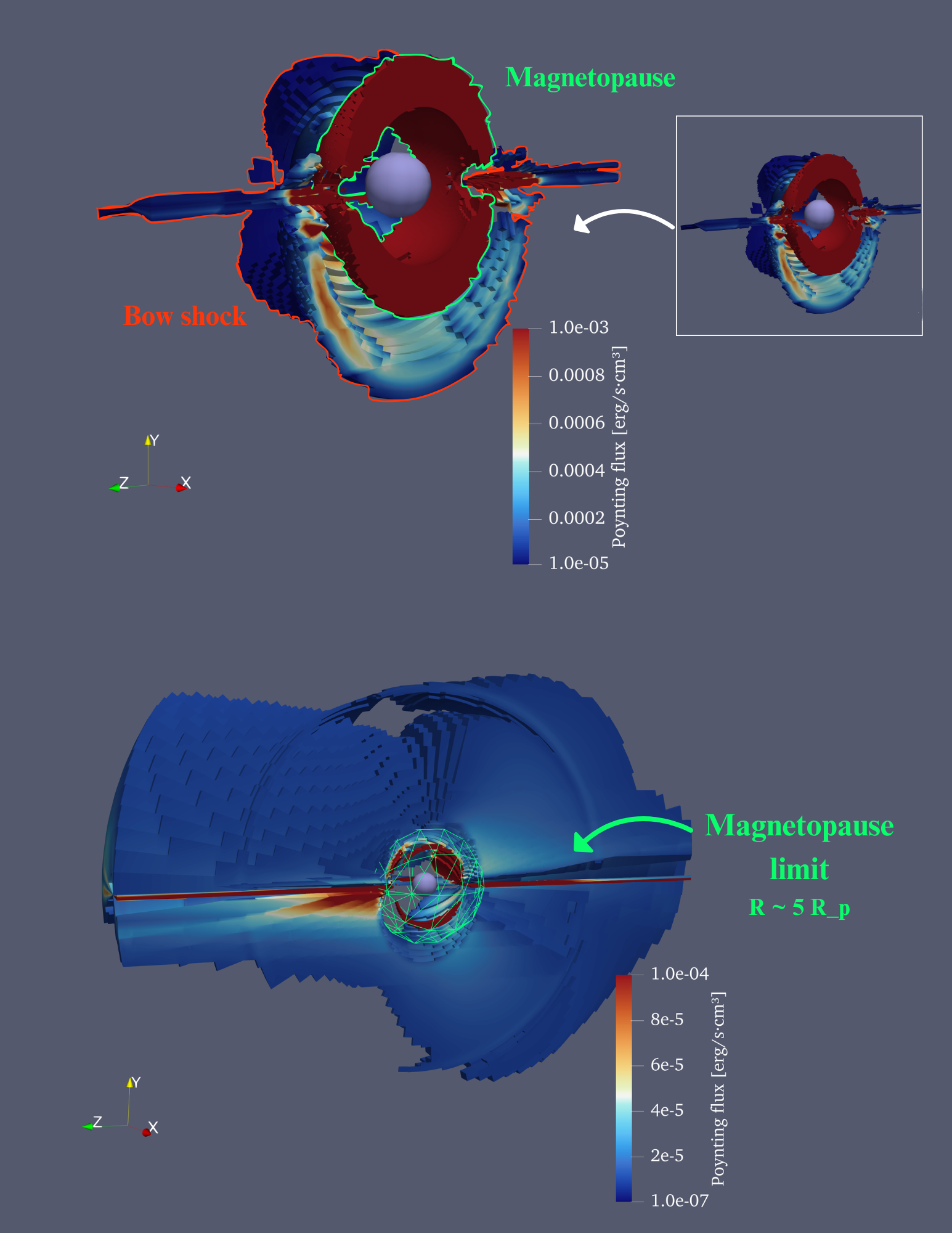}
\end{subfigure}
\caption{Contributions to the peak $P_{R}$ from the bow shock and magnetopause at $\eta=5.38018\times10^{8}$~cm$^{2}$~s$^{-1}$ for Case~\#1 (panel~a), Case~\#2 (top of panel~b), and Case~\#3 (bottom of panel~b). The volumetric renderings show $\nabla\cdot\mathbf{S}_{\rm total}$ in erg~cm$^{-3}$~s$^{-1}$.}
\label{fig:PR_estimations}
\end{figure*}

We estimate $P_{R}$ in the bow shock and magnetopause using equation~(\ref{eq:PR_def}), with $P_{B}$ given by equation~(\ref{eq:PB_total}) for all cases. We integrate the emission proxy over the dayside interaction volume ($X=2$ to $X\sim-10~R_p$), including both the bow shock and magnetopause. In Case~\#1, these regions are clearly separated (Figure~\ref{fig:PR_estimations}a), and we integrate each contribution directly. In the more compressed Cases~\#2 and \#3 (Figure~\ref{fig:PR_estimations}b), the separation is less visible; we therefore partition the volume using a radial cut at $R\simeq5~R_p$ and compute each contribution by excluding the complementary region. Because this partition is not fully robust when the layers overlap, we regard the total integrated emission as the most reliable estimate. Figure~\ref{fig:PR_estimations} highlights the dominant emitting regions via volumetric renderings of $\nabla\cdot\mathbf{S}_{\rm total}$ (erg~cm$^{-3}$~s$^{-1}$): thin, well-defined bow-shock and magnetopause layers in Case~\#1, and a more compact, asymmetric volume in Cases~\#2--\#3, bounded by a magnetopause limit near $R\sim5R_p$.

Even in the compressed cases, $P_{R}$ remains concentrated in two layers: an outer arc associated with the bow shock and an inner arc associated with the magnetopause, where field-line draping and flow diversion occur. The regional budgets in Table~\ref{tab:PR_emission_eta} use masks tied to these boundaries (outer shocked layer versus inner coupling layer). We caution, however, that when the layers become thin and partially overlap, reconnection-driven broadening can smear $P_{R}$ across the magnetosheath, introducing ambiguity in the bow-shock versus magnetopause partition, whereas the total power integrated over both layers remains comparatively robust.

Table~\ref{tab:PR_emission_eta} reports the radio-power proxy $P_{R}$ inferred from equation~(\ref{eq:PB_total}), separated into bow-shock and magnetopause regions and their sum. We emphasize that this quantity represents a net Poynting-flux budget over the chosen volume.

For Cases~\#1--\#2, $P_{R}$ is $\sim10^{19}$--$10^{20}$~$\mathrm{erg~s^{-1}}$ at $\eta=0$--$538.018$~$\mathrm{cm^{2}~s^{-1}}$, increases to $\sim10^{24}$--$10^{26}$~$\mathrm{erg~s^{-1}}$ for $\eta=5.38018\times10^{8}$--$5.38018\times10^{12}$~$\mathrm{cm^{2}~s^{-1}}$. In the imposed-$\eta$ regime, the magnetopause contribution becomes comparable to or larger than the bow-shock term in Cases~\#1--\#2, consistent with a redistribution of the net energy-conversion toward the inner coupling layer as current sheets broaden. Case~\#3 is systematically higher at low $\eta$ (total $\sim10^{22}$--$10^{23}$~$\mathrm{erg~s^{-1}}$ for $\eta=0$--$5.38018\times10^{8}$~$\mathrm{cm^{2}~s^{-1}}$) and rises to $\sim10^{25}$~$\mathrm{erg~s^{-1}}$ at $\eta=5.38018\times10^{12}$~$\mathrm{cm^{2}~s^{-1}}$.

\begin{table}
\centering
\caption{Emitted power from the bow shock, the magnetopause, and their sum (total) for all simulation cases.}
\label{tab:PR_emission_eta}
\begin{tabularx}{\columnwidth}{l *{3}{>{\centering\arraybackslash}X}}
\hline
Case ($\eta$ in cm$^{2}$ s$^{-1}$) & Bow shock (erg s$^{-1}$) & Magnetopause (erg s$^{-1}$) & Total emission (erg s$^{-1}$) \\
\hline
\#1 ($\eta=0$) & $3.34\times10^{19}$ & $6.15\times10^{18}$ & $3.95\times10^{19}$ \\
\#1 ($\eta=538.018$) & $4.16\times10^{19}$ & $3.52\times 10^{19}$ & $7.68\times 10^{19}$  \\
\#1 ($\eta=5.38018\times10^{8}$) & $5.57\times 10^{24}$ & $5.74\times 10^{23}$ & $6.14\times 10^{24}$  \\
\#1 ($\eta=5.38018\times10^{12}$) & $9.57\times 10^{25}$ &  $3.45\times 10^{24}$ & $9.91\times 10^{25}$   \\
\#2 ($\eta=0$) & $2.81\times10^{19}$ & $2.39\times10^{19}$ & $5.20\times10^{19}$ \\
\#2 ($\eta=538.018$) & $6.48\times10^{19}$ & $2.33\times10^{20}$ & $2.97\times10^{20}$ \\
\#2 ($\eta=5.38018\times10^{8}$) & $3.97\times10^{21}$ &  $1.97\times10^{23}$ & $2.00\times10^{23}$ \\
\#2 ($\eta=5.38018\times10^{12}$) & $6.86\times10^{24}$ & $2.47\times10^{24}$ & $9.33\times10^{24}$  \\
\#3 ($\eta=0$) & $8.16\times10^{21}$ & $4.64\times10^{21}$ & $1.28\times10^{22}$ \\
\#3 ($\eta=538.018$) & $7.28\times10^{22}$ & $9.20\times10^{20}$ & $7.37\times10^{22}$ \\
\#3 ($\eta=5.38018\times10^{8}$) & $8.10\times10^{22}$ & $6.05\times10^{22}$ & $1.41\times10^{23}$  \\
\#3 ($\eta=5.38018\times10^{12}$) & $9.23\times10^{24}$ & $4.41\times10^{24}$ & $1.36\times10^{25}$ \\
\hline
\end{tabularx}
\end{table}

\begin{figure*}
\centering
\includegraphics[width=18.0cm, height=6.0cm]{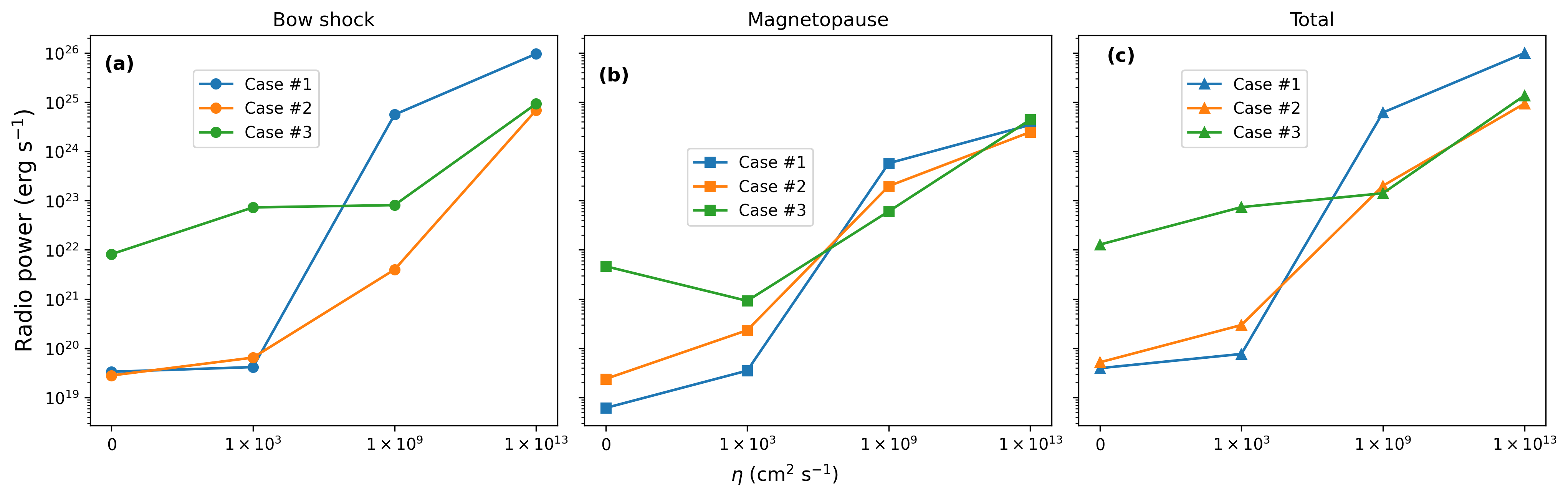}
\caption{Integrated radio-power proxy in the bow shock (panel~a), magnetopause (panel~b), and total (panel~c) for Cases~\#1--\#3 as a function of the four magnetic diffusivity values. Both axes are logarithmic.}
\label{fig:radio_power_vs_res}
\end{figure*}

Figure~\ref{fig:radio_power_vs_res} shows the integrated radio-power proxy as a function of magnetic diffusivity for the three interaction cases, separated into the bow-shock contribution (panel a), the magnetopause contribution (panel b), and their sum (panel c). In all cases, the total emitted power increases with $\eta$, although the rate of increase differs among the three regimes. Case~\#1 exhibits the strongest growth, especially in the bow-shock contribution, which becomes dominant at high diffusivity. Case~\#2 shows a more gradual rise, with comparable bow-shock and magnetopause contributions at the largest $\eta$. Case~\#3 starts from a systematically higher power at low diffusivity and remains comparatively strong across the full range, although its increase is less abrupt than in Case~\#1. Overall, the figure indicates that increasing magnetic diffusivity enhances the electromagnetic power budget inferred for radio emission while also modifying the relative importance of the bow shock and magnetopause in each case.

In this work, the expression ``efficient energy transport'' is used in a qualitative, geometrical sense and should not be interpreted as a lossless transmission of electromagnetic energy toward the exoplanet. Our diagnostics do not measure transport efficiency in a strict sense; rather, we quantify how the incident Poynting flux carried by the stellar wind is redistributed and locally converted within the dayside interaction region. In particular, regions with positive $\nabla\cdot\mathbf{S}_{\rm total}$ identify sites of electromagnetic energy conversion within the resistive-MHD framework. Variations in magnetic diffusivity modify the topology and thickness of the coupling layer and shift the spatial location of these conversion regions. Therefore, $P_{R}$ reflects the total electromagnetic energy conversion within the selected volumes, rather than being an intrinsic measure of how efficiently energy is transported inward without losses.

\begin{figure*}
\begin{subfigure}[b]{0.3\textwidth}
\centering
\caption{}
\includegraphics[width=\columnwidth]{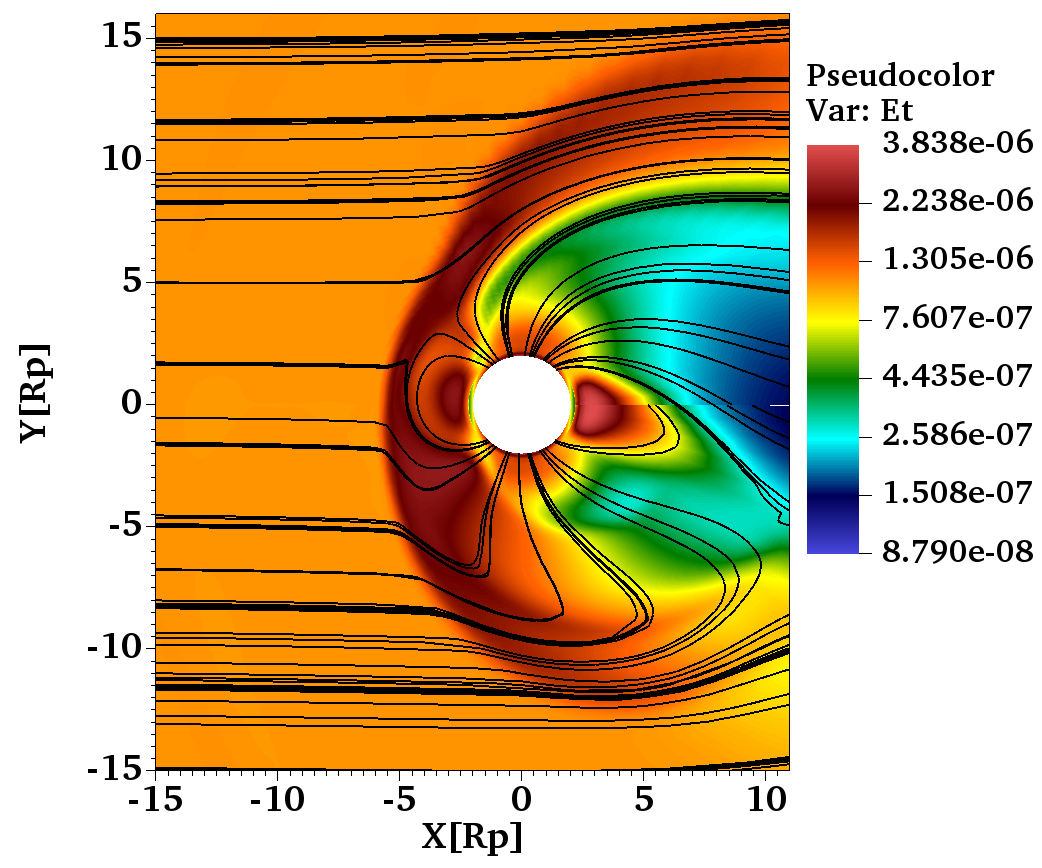}
\end{subfigure}
\begin{subfigure}[b]{0.3\textwidth}
\centering
\caption{}
\includegraphics[width=\columnwidth]{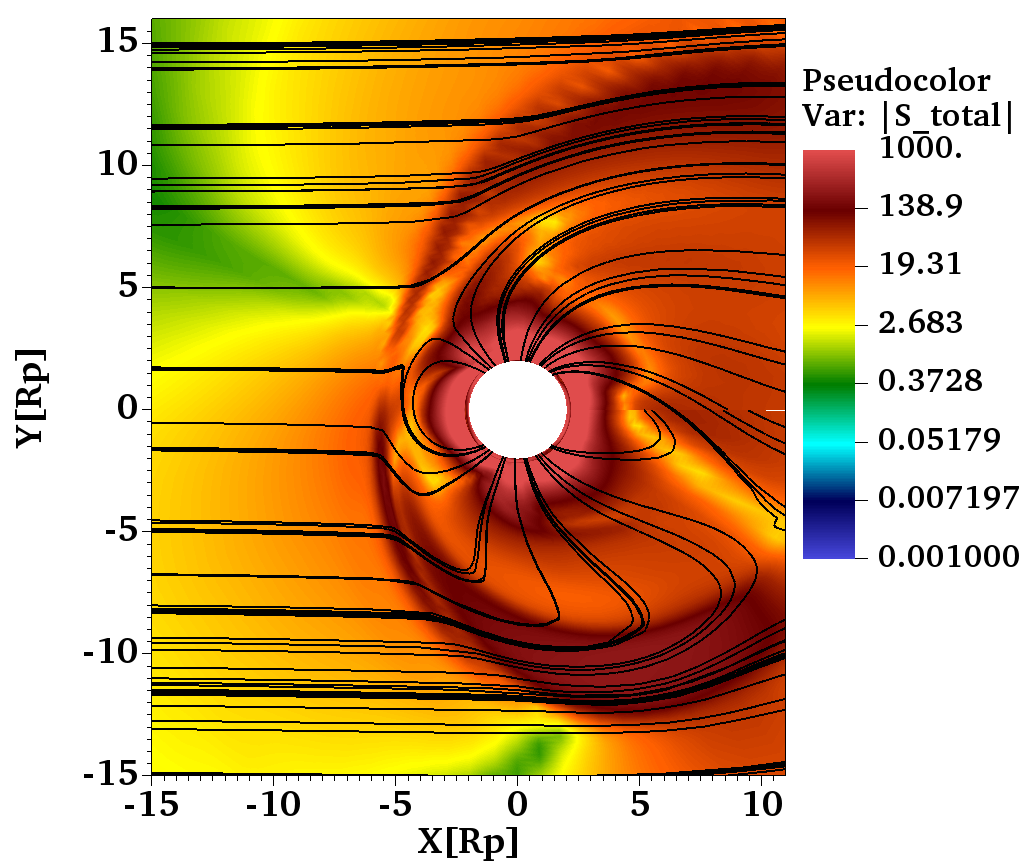}
\end{subfigure}
\begin{subfigure}[b]{0.3\textwidth}
\centering
\caption{}
\includegraphics[width=\columnwidth]{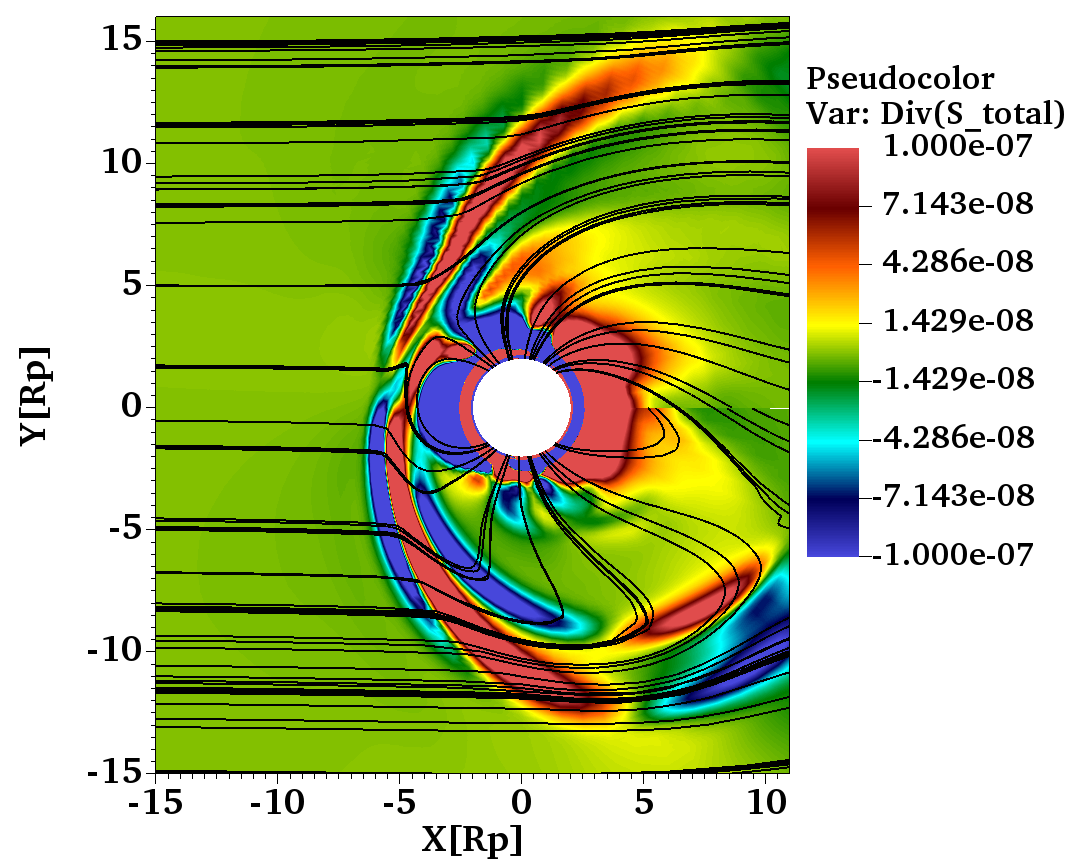}
\end{subfigure}
\begin{subfigure}[b]{0.3\textwidth}
\centering
\caption{}
\includegraphics[width=\columnwidth]{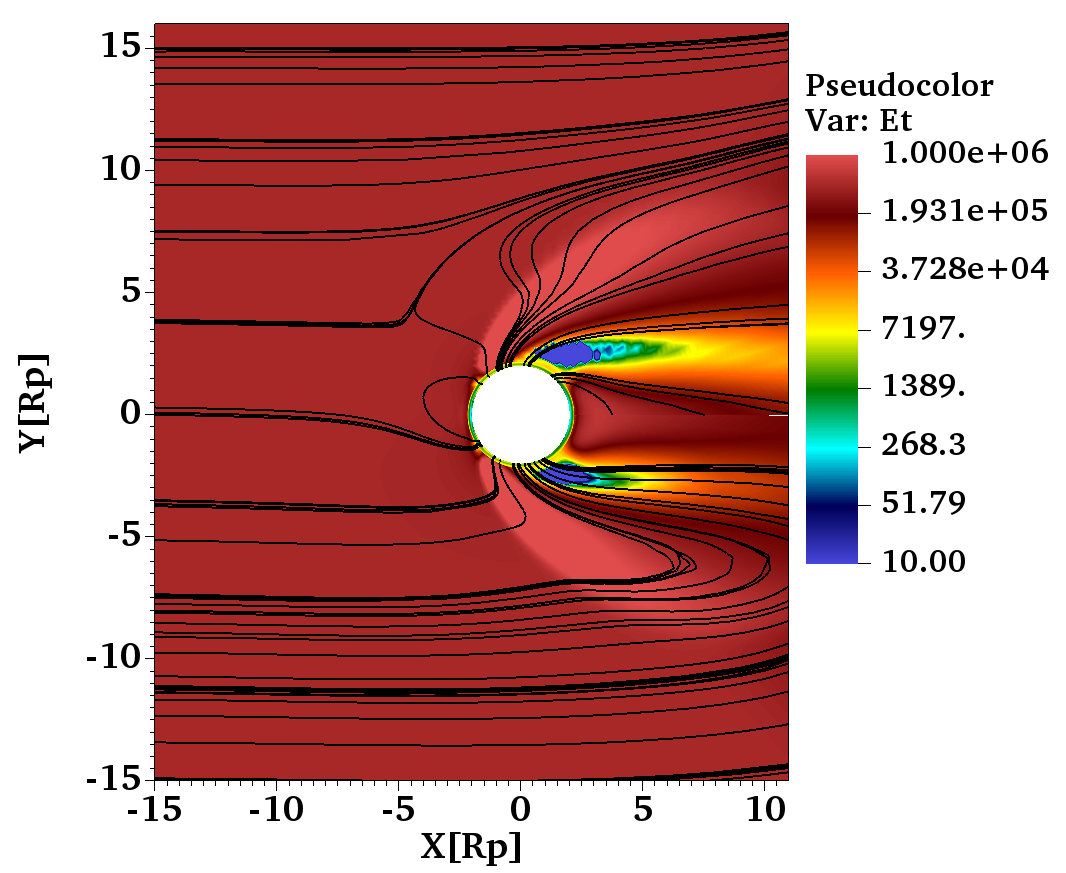}
\end{subfigure}
\begin{subfigure}[b]{0.3\textwidth}
\centering
\caption{}
\includegraphics[width=\columnwidth]{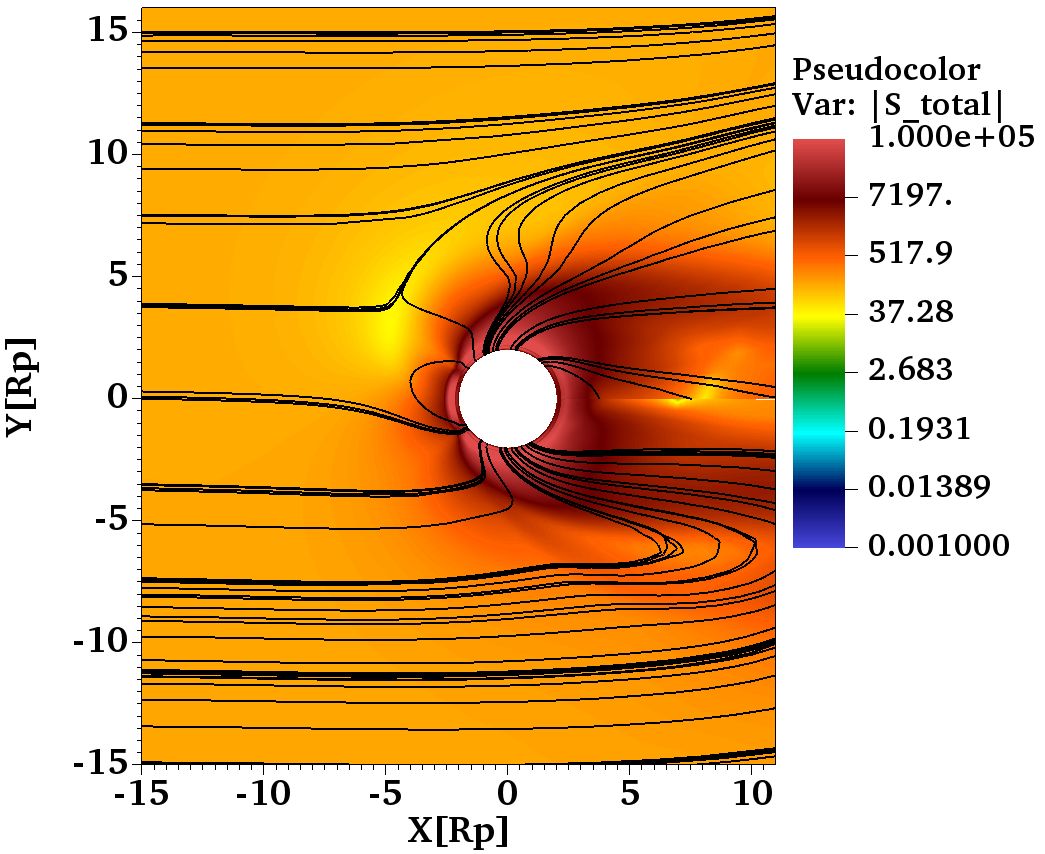}
\end{subfigure}
\begin{subfigure}[b]{0.3\textwidth}
\centering
\caption{}
\includegraphics[width=\columnwidth]{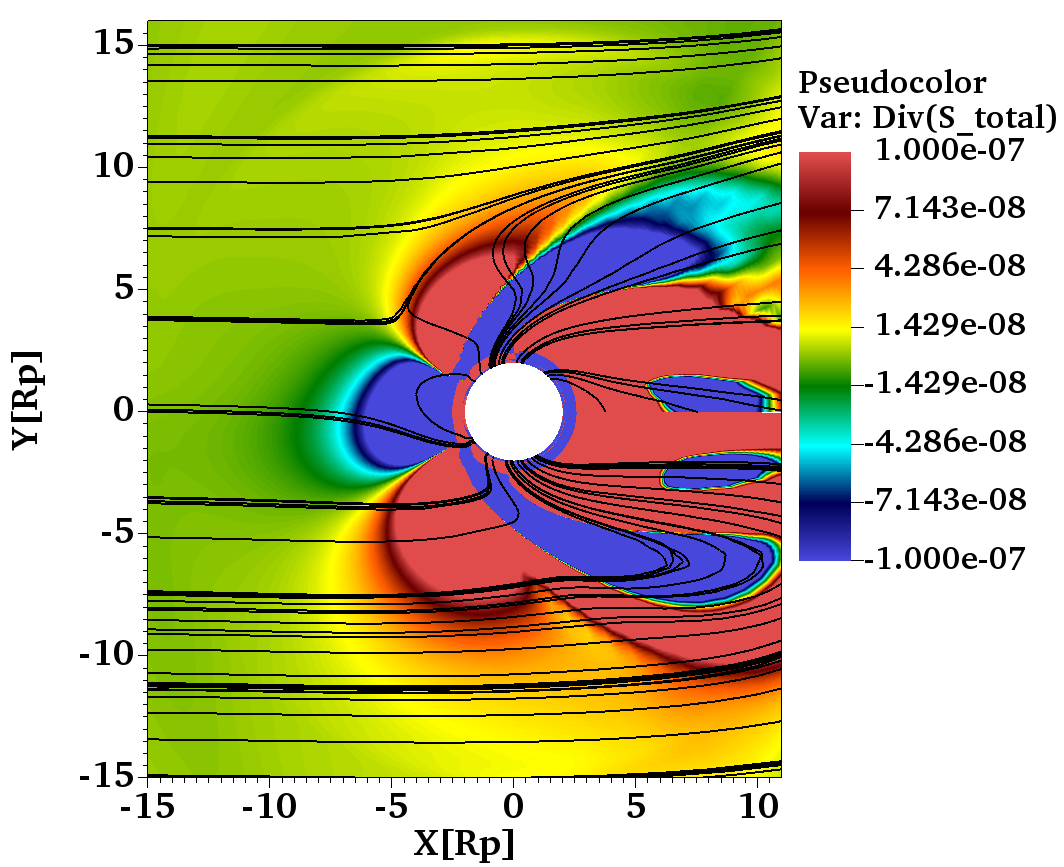}
\end{subfigure}
\begin{subfigure}[b]{0.3\textwidth}
\centering
\caption{}
\includegraphics[width=\columnwidth]{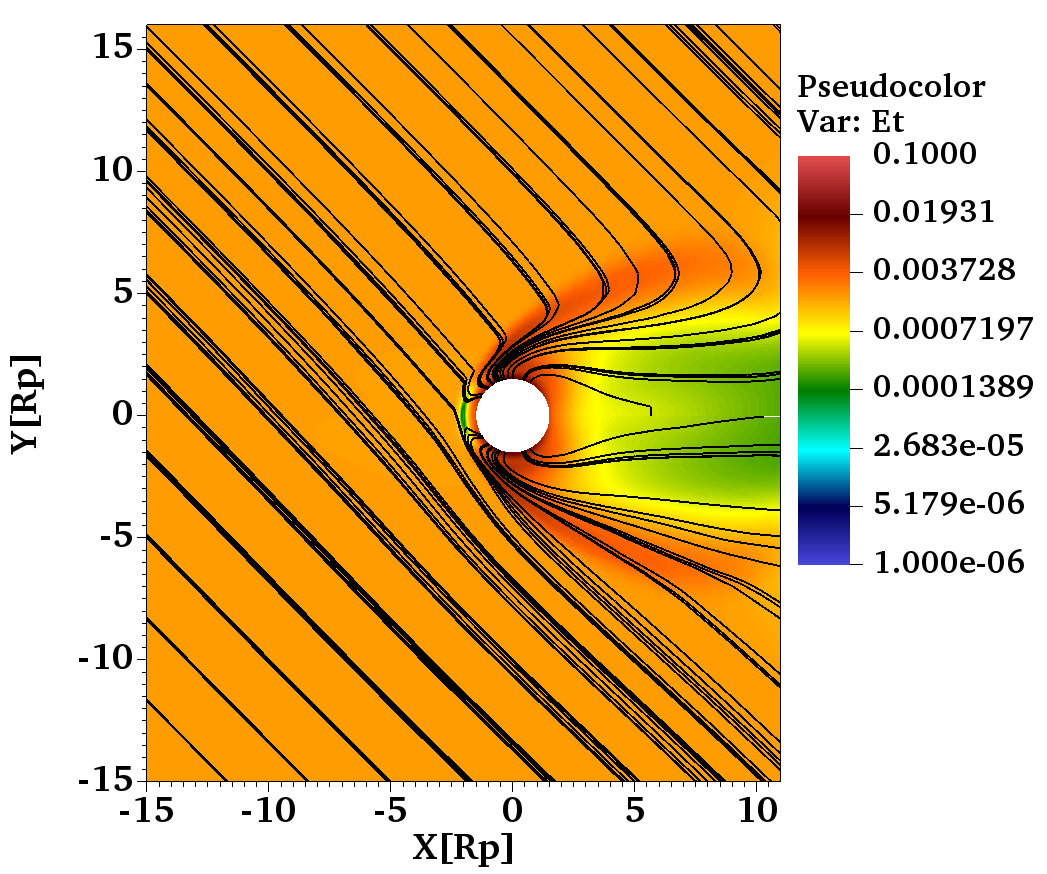}
\end{subfigure}
\begin{subfigure}[b]{0.3\textwidth}
\centering
\caption{}
\includegraphics[width=\columnwidth]{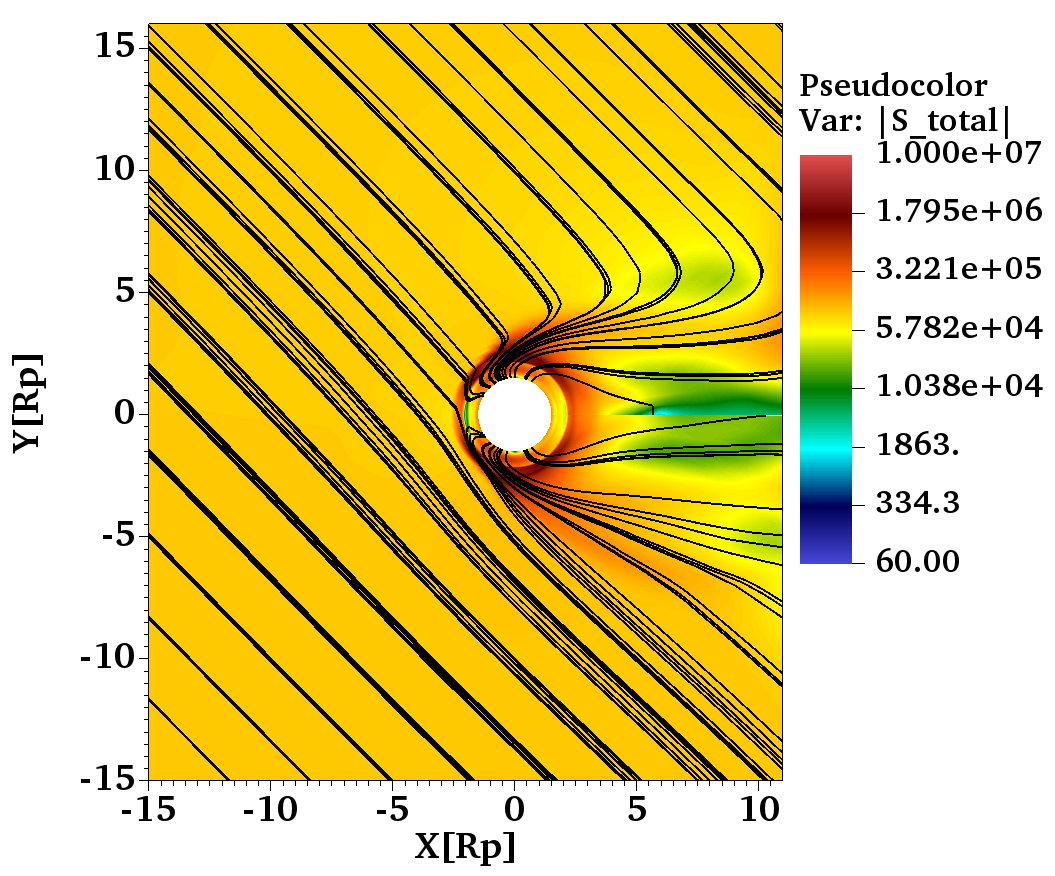}
\end{subfigure}
\begin{subfigure}[b]{0.3\textwidth}
\centering
\caption{}
\includegraphics[width=\columnwidth]{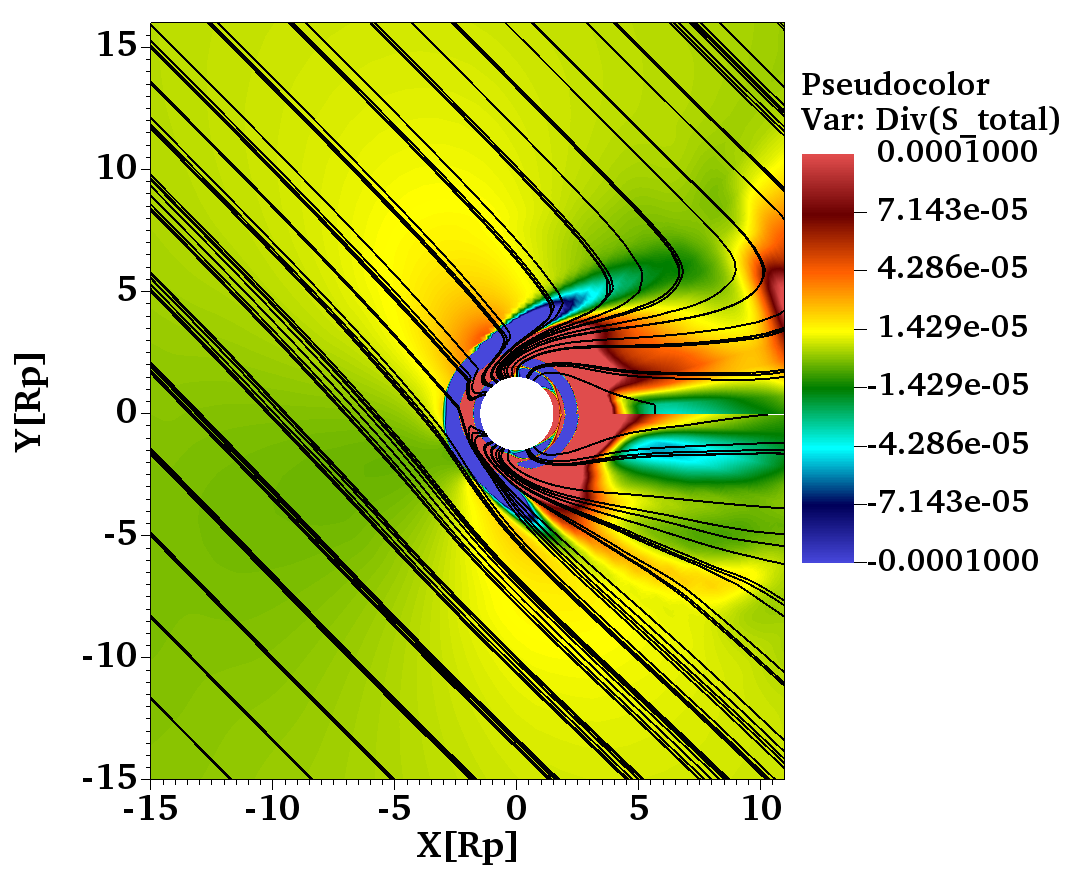}
\end{subfigure}
\caption{Equatorial ($x$-$y$) maps for $\eta=0$ showing the logarithm of the total energy density $E_t$ (panels~a, d, g; erg~cm$^{-3}$), the logarithm of the total Poynting-flux magnitude $|\mathbf{S}_{\rm total}|$ (panels~b, e, h; erg~cm$^{-2}$~s$^{-1}$), and $\nabla\!\cdot\!\mathbf{S}_{\rm total}$ (panels~c, f, i; erg~cm$^{-3}$~s$^{-1}$). Rows correspond to Case~\#1 (top), Case~\#2 (middle), and Case~\#3 (bottom). Black streamlines trace magnetic field lines.}
\label{fig:Results_flux_eta_0}
\end{figure*}

Figure~\ref{fig:Results_flux_eta_0} summarizes dayside energy transport in the equatorial ($x$-$y$) plane for $\eta=0$. The left column shows the logarithm of the total energy density $E_t$ (panels~a, d, and g), defined here as the sum of kinetic, thermal, and magnetic energy densities, and thus indicates where stellar wind energy accumulates in the compressed interaction region. The middle column shows the logarithm of the total Poynting-flux magnitude $|\mathbf{S}_{\rm total}|$ (panels b, e, and h), which is helpful for tracing the local electromagnetic energy flux and highlights the dominant transport pathways. The right column shows $\nabla\cdot\mathbf{S}_{\rm total}$ (panels~c, f, and i), the local divergence of the electromagnetic energy flux. Negative $\nabla\cdot\mathbf{S}_{\rm total}$ indicates Poynting-flux convergence (net electromagnetic energy entering the local volume and being transferred through work on the plasma and diffusivity effects), whereas positive values indicate net divergence (net electromagnetic energy leaving the local volume). Furthermore, the magnetic field lines schematize a draping and connectivity guide for the Poynting flux.

In Case~\#1 (top row), $E_t$ and $|\mathbf{S}_{\rm total}|$ outline a well-developed bow shock and magnetosheath. The bow shock is the outer compressive layer where the stellar wind is decelerated and heated, and the magnetopause is the inner boundary where the draped stellar-wind field and the exoplanetary obstacle produce strong shear. The $\nabla\cdot\mathbf{S}_{\rm total}$ map shows alternating regions of convergence and divergence concentrated in thin arcs near these boundaries, consistent with electromagnetic energy conversion localized to the bow-shock and magnetopause layers. In Case~\#2 (middle row), the interaction is more strongly compressed, bringing the bow shock and magnetopause closer to the exoplanet and narrowing the downstream channel. The enhanced $E_t$ and $|\mathbf{S}_{\rm total}|$ are therefore confined to a smaller dayside region, and the $\nabla\cdot\mathbf{S}_{\rm total}$ pattern becomes more concentrated near the coupling layer, reflecting a thinner sheath and reduced spatial separation between the outer shocked layer. In Case~\#3 (bottom row), the interaction region is the most compact, and the wake is most collimated. The maxima of $E_t$ and $|\mathbf{S}_{\rm total}|$ occur close to the dayside boundary and along the narrow downstream channel, while $\nabla\cdot\mathbf{S}_{\rm total}$ remains localized near the dayside coupling layer, where the strongest gradients in $\mathbf{B}$ and $\mathbf{v}$ are visible.

In general, the three cases show a similar physical picture: stellar wind energy accumulates in the dayside compression region (high $E_t$), electromagnetic energy transport is guided around the exoplanet by draped field lines (enhanced $|\mathbf{S}_{\rm total}|$ along the magnetopause and into the wake), and the dominant electromagnetic energy conversion occurs in thin boundary layers (largest $|\nabla\cdot\mathbf{S}_{\rm total}|$) associated with the bow shock and magnetopause layers. The primary difference among the cases is the degree of dayside compression, which increases from Case~\#1 to Cases~\#2 and \#3, moving these layers closer to the exoplanet, narrowing the wake, and reducing the separation between the outer shocked layer and the inner coupling layer.

\section{Discussion and Conclusions}
\label{Discussion_and_conclusions}

We performed global three-dimensional resistive MHD simulations of the TRAPPIST-1 wind interacting with a dipolar TRAPPIST-1e magnetosphere in three forcing regimes (Cases~\#1--\#3), scanning four spatially constant magnetic diffusivities from $\eta=0$ to $5.38\times10^{12}$~$\mathrm{cm^{2}~s^{-1}}$. Electromagnetic energy transport is diagnosed using maps of the total energy density $E_t$, the total Poynting-flux magnitude $|\mathbf{S}_{\rm total}|$, and $\nabla\cdot\mathbf{S}_{\rm total}$ in the dayside interaction region. As a proxy for the electromagnetic power available for radio emission, we post-process the simulations using the signed volume integral of $\nabla\cdot\mathbf{S}_{\rm total}$ over a dayside interaction volume that includes the bow shock and magnetopause (with a case-dependent geometric partition when needed). In resistive MHD, $\mathbf{S}_{\rm total}=\mathbf{S}_{\rm ideal}+\mathbf{S}_{\rm res}$, where $\mathbf{S}_{\rm res}\propto \eta(\nabla\times\mathbf{B})\times\mathbf{B}$; this makes the proxy explicitly sensitive to $\eta$ through both the solution and the energy-flux definition. We adopt the radio--magnetic scaling $P_R=\beta P_B$ with $\beta\sim2\times10^{-3}$ \citep{Zarka2018}.

\paragraph*{Main physical trends and novelty.}
Across all cases, increasing $\eta$ broadens the dayside coupling layer and relaxes sharp current structures, so the electromagnetic energy-conversion transitions from thin, patchy filaments at low $\eta$ to thicker, more spatially extended arcs and patches at high $\eta$. This redistribution is based on the following: in the $\nabla\cdot\mathbf{S}_{\rm total}$ budget, $P_{R}$ spans many orders of magnitude across $\eta$, and the relative weight of the bow-shock and magnetopause contributions can change substantially. A key novelty of this work is to demonstrate, within a controlled global setup, how non-ideal coupling modifies (i) the spatial organization of dayside energy transport (where the conversion is concentrated) and (ii) the inferred radio-power (how large the net electromagnetic budget can become) when $\eta$ is varied.

\paragraph*{Constraints of physical diffusivity in terms of numerical diffusivity}
For the present grid, HLL solver, and minmod reconstruction, we estimate an effective numerical diffusivity of order $\eta_{\rm num}\sim10^{15}$--$10^{16}$ cm$^{2}$ s$^{-1}$ in the strongest-gradient regions. Since all explicitly imposed values in our scan, including the largest one ($\eta = 5.38018\times10^{12}$ cm$^{2}$ s$^{-1}$), remain below this estimate, the present calculations should not be interpreted as providing a direct measurement of the physical diffusivity of the TRAPPIST-1e environment. Instead, they define a controlled set of prescribed transport coefficients that are useful for identifying robust trends in the global wind--magnetosphere coupling and in the radio-power proxy. Any attempt to translate the present $\eta$ values into physical diffusivities therefore requires caution and should be supported by resolution studies, alternative numerical schemes, or extended models that include additional non-ideal effects such as Hall or kinetic physics.

\paragraph*{What the radio-power proxy does and does not measure.}
Our proxy is a net Poynting-flux budget over a prescribed dayside volume (equation~\ref{eq:PB_total}), not a strictly positive-definite local dissipation rate. In particular, $\nabla\cdot\mathbf{S}_{\rm total}$ can be positive or negative locally, and the signed integral measures the net electromagnetic energy conversion within the chosen region rather than the total irreversible heating. This is appropriate for an estimate of the ``power available for radio emission'' in the radio--magnetic spirit, but it implies that the partition between subregions (bow shock versus magnetopause masks) is intrinsically less robust than the total integrated budget when the layers overlap. Future work could complement the present diagnostic with an explicitly positive-definite dissipation proxy (for example, $\eta J^2$ in resistive MHD, when available in post-processing) to disentangle net electromagnetic redistribution from irreversible conversion.

\paragraph*{Detectability and observational frequency constraints.}
A separate and decisive constraint is the emission frequency. For cyclotron-maser emission, the maximum observable frequency is set by the local electron cyclotron frequency,
\begin{equation}
\nu_{\rm c}\simeq \frac{eB}{2\pi m_e c}\approx 2.8~B~\mathrm{MHz},
\end{equation}
with $B$ in gauss. Using the exoplanetary dipole strengths adopted in our three cases, $B_{\rm TRAPPIST-1e}=0.32,\ 0.64,\ 1.28$~$\mathrm{G}$, the corresponding values are $B_{\rm pole}\simeq2B_{\rm eq}$ for a dipole, giving $\nu_{c,\max}\approx 5.6~B_{\rm eq}~\mathrm{MHz}\approx {1.79,\ 3.58,\ 7.17}~\mathrm{MHz}$ for Cases~\#1, \#2, and \#3. These cutoff frequencies lie below the terrestrial ionospheric transparency limit, since radiation below about $10$~$\mathrm{MHz}$ does not penetrate the ionosphere and is therefore inaccessible to ground-based facilities. Moreover, current major low-frequency ground arrays operate at substantially higher frequencies; for example, LOFAR at roughly $10$--$240$~$\mathrm{MHz}$ \citep{vanHaarlem2013} and SKA1-Low at $50$--$350$~$\mathrm{MHz}$ \citep{SKAOScienceTeam2015}, well above the predicted $\nu_{\rm cmax}$. In this parameter set, a detection of TRAPPIST-1e cyclotron emission from the ground is therefore not expected. A ground-accessible cutoff at $\nu_{\rm cmax}\gtrsim10$~$\mathrm{MHz}$ would require $B_{\rm pole}\gtrsim3.6$~$\mathrm{G}$, i.e., $B_{\rm eq}\gtrsim1.8$~$\mathrm{G}$, which is above our assumed exoplanetary fields.

Space-based low-frequency concepts, including lunar-farside arrays designed to operate in the sub-10~MHz regime, would be better matched to the predicted cutoffs. For example, FARSIDE targets approximately $0.1$--$40$~$\mathrm{MHz}$ \citep{Valinia2022}. Even in that case, detectability would still depend on uncertain factors not captured by our proxy $P_{R}$, including the emission beaming pattern, the conversion efficiency $\beta$, propagation and absorption in the stellar environment, and contamination by stellar coherent bursts. Therefore, our main observational conclusion is clear: with the exoplanetary field strengths adopted here, the relevant emission frequencies are likely below the ground-based window, so constraining the TRAPPIST-1e magnetosphere through radio observations requires space-based low-frequency capability or substantially stronger exoplanetary fields than those assumed.

\paragraph*{Implications of magnetic diffusivity for future studies and missions.}
The strong sensitivity of the radio-power proxy $P_{\rm R}$ to $\eta$ has direct consequences for how simulation outputs should be compared with observational data. In our framework, $\eta$ controls the topology of the dayside coupling region, which in turn regulates where $\nabla \cdot \mathbf{S}_{\rm total}$ concentrates and how much net electromagnetic energy enters the interaction volume. Across the imposed-diffusivity scan, the energy-conversion proxy becomes less localized, and the magnetopause contribution can grow to match or exceed the bow-shock contribution. Therefore, if the effective diffusivity is overestimated, the model can predict broader conversion regions and larger integrated budgets than would occur in a less diffusive plasma, potentially biasing the inferred radio power upward. Conversely, if the effective physical diffusivity is lower than assumed, energy conversion may remain more confined to thin layers, and the total proxy may be smaller, even if the large-scale magnetospheric geometry is similar. This sensitivity implies that comparisons between predicted powers and observational upper limits should not be treated as one-to-one constraints on the exoplanetary magnetic field without accounting for uncertainty in the effective diffusivity and emission efficiency. A non-detection can be consistent with a magnetosphere that is present but less diffusive, such that the net conversion proxy and the associated radio power are reduced, or with emission that is shifted in frequency and beaming. For this reason, we interpret our $\eta$ scan as a controlled sensitivity study and emphasize trends that are robust across regimes. Future radio-oriented studies should therefore bracket predictions by exploring a wider range of effective transport prescriptions, together with at least one resolution check, and should report observational comparisons in terms of ranges or envelopes in $P_{\rm R}$ rather than single values. This is particularly relevant for next-generation low-frequency missions, for which a reliable translation from an observed upper limit into a constraint on magnetospheric properties will require modeling that ties $\eta$ to reconnection physics or turbulence, or uses Hall MHD or kinetic extensions calibrated against Solar System benchmarks.

\paragraph*{Model limitations.} 
An important caveat is that our results are obtained within a single-fluid, resistive-MHD framework with an explicitly prescribed magnetic diffusivity that is isotropic and spatially uniform. This approach captures global, system-scale wind--magnetosphere coupling and provides a self-consistent way to quantify energy-transport and conversion proxies within MHD, but it does not include Hall or kinetic physics that can control reconnection, current-sheet microstructure, species-dependent heating, and non-thermal particle acceleration. As a result, the reconnection rates, the thickness and localization of current layers, and the partition of converted magnetic energy into thermal versus kinetic channels may differ from those in the actual TRAPPIST-1e environment. In addition, because our radio-power estimate relies on a Poynting-flux budget proxy rather than a positive-definite dissipation rate, any mapping from the integrated proxy to auroral or radio emission should be interpreted cautiously in terms of absolute magnitude. We estimate an effective numerical magnetic diffusivity of order $\eta_{\rm num}\sim10^{15}$--$10^{16}$ cm$^{2}$ s$^{-1}$ in the strongest-gradient regions, whereas all explicitly imposed values in the present scan, including the largest one, remain below this range. The present $\eta$ scan should therefore be interpreted primarily as a controlled sensitivity study, useful for identifying robust trends in the global wind--magnetosphere coupling and in the radio-power proxy, rather than as a direct measurement of the physical diffusivity of the TRAPPIST-1e environment. Any quantitative translation from the imposed $\eta$ values to physical transport coefficients will require resolution studies, alternative numerical schemes, or extended models that include additional non-ideal effects such as Hall or kinetic physics.

\paragraph*{Summary.}
Within a global resistive-MHD framework, we find that magnetic diffusivity regulates the thickness and topology of the dayside coupling layer and strongly influences the integrated Poynting-flux budget proxy $P_{R}$ used to estimate radio power. Interpreting these trends requires explicit accounting for the numerical diffusivity floor. For the exoplanetary field strengths explored here, the expected cyclotron cutoff frequencies are $\lesssim 7.2$~$\mathrm{MHz}$, implying that direct ground-based radio detection is unlikely and that meaningful radio constraints on the magnetosphere will require space-based observations below 10~MHz or substantially stronger exoplanetary fields than assumed. In future work, we will extend this framework to Solar System magnetospheres (Earth and Mercury), where in situ measurements enable stringent tests of boundary locations, reconnection signatures, and global energy transfer. We will also explore spatially non-uniform, turbulence-driven diffusivity prescriptions, assess Hall and ambipolar effects, and couple the global MHD solutions to kinetic models of particle acceleration to produce predictions of dynamic radio spectra and polarization.

\section*{Acknowledgements}
J.J.G.A.\ acknowledges support from the UNAM--PAPIIT grant IA100725, and A.S.\ acknowledges support from the UNAM--PAPIIT grant IN115524. The authors thank Miljenko \v{C}emelji\'c and Jacobo Varela for providing the PLUTO code template for the star--exoplanet interaction setup. We also thank Jacobo Varela for a constructive report that substantially improved the manuscript.

\section*{Data Availability}

The data underlying this article will be shared on reasonable request to the corresponding author.
 



\bibliographystyle{mnras}
\bibliography{Resistive_MHD_T1e_MNRAS_2026} 








\bsp	
\label{lastpage}
\end{document}